\documentclass[twoside,leqno]{article}
\usepackage{siamproceedings}
\usepackage{color, xcolor, colortbl}
\usepackage{graphicx}
\usepackage[letterpaper,top=1.5in]{geometry}
\usepackage{amsfonts}
\usepackage{algorithm}
\usepackage{algorithmic}
\usepackage{dcolumn}
\usepackage{epstopdf}
\usepackage{bm}
\usepackage[caption=false]{subfig}
\usepackage{appendix}
\usepackage{multirow}
\usepackage{braket}
\usepackage[english]{babel}
\usepackage{hyperref}
\usepackage[capitalize]{cleveref}
\crefname{section}{Section}{Sections}
\crefname{subsection}{Section}{Sections}
\Crefname{section}{Section}{Sections}
\Crefname{subsection}{Section}{Sections}

\usepackage[square,numbers]{natbib}
\usepackage[T1]{fontenc}
\usepackage{xpatch}
\usepackage{tikz}
\usepackage{adjustbox}
\usepackage{xspace}
\usepackage[roman]{complexity}
\usepackage{xparse}
\usepackage{orcidlink}

\newcommand{\bvec}[1]{\mathbf{#1}}

\newcommand{\va}{\bvec{a}}
\newcommand{\vb}{\bvec{b}}
\newcommand{\vc}{\bvec{c}}

\newcommand{\ve}{\bvec{e}}

\newcommand{\vx}{\bvec{x}}

\renewcommand{\Re}{\operatorname{Re}}
\renewcommand{\Im}{\operatorname{Im}}

\newcommand{\I}{i}

\newcommand{\wt}[1]{\widetilde{#1}}

\newcommand{\abs}[1]{\left\lvert#1\right\rvert}
\newcommand{\norm}[1]{\left\lVert#1\right\rVert}

\newcommand{\ud}{\,\mathrm{d}}
\newcommand{\Or}{\mathcal{O}}

\newcommand{\NN}{\mathbb{N}}
\newcommand{\RR}{\mathbb{R}}
\renewcommand{\CC}{\mathbb{C}}
\newcommand{\ZZ}{\mathbb{Z}}

\NewDocumentCommand{\ketbra}{mG{#1}}{\mathinner{|{#1}\rangle\!\langle{#2}|}}

\usetikzlibrary{quantikz2}

\usetikzlibrary{fit}
\tikzset{%
  highlight/.style={rectangle,rounded corners,fill=blue!15,draw,fill opacity=0.3,thick,inner sep=0pt}
}

\usepackage{enumerate}
\usepackage{enumitem}
\newcommand{\argmin}{\mathop{\mathrm{argmin}}}
\newcommand{\ol}[1]{\ensuremath{\overline{#1}}}
\newcommand{\lzero}{\ell_0 (\ZZ)}

\newcommand{\ba}{\mathbf{a}}
\newcommand{\bb}{\mathbf{b}}

\newcommand{\bga}{\boldsymbol{\gamma}}

\newcommand{\DD}{\mathbb{D}}
\newcommand{\TT}{\mathbb{T}}

\newcommand{\cDD}{\overline{\mathbb{D}}}
\newcommand{\ceil}[1]{\left\lceil #1\right\rceil}

\DeclareMathOperator*{\esssup}{\mathrm{ess\,sup}}
\DeclareMathOperator{\Arg}{Arg}

\begin{document}

\title{\Large Mathematical and numerical analysis of quantum signal processing\thanks{Submitted to the Proceedings of the 2026 International Congress of Mathematicians (ICM 2026).}}
\author{Lin Lin\thanks{Department of Mathematics, University of California, Berkeley; Applied Mathematics and Computational Research Division, Lawrence Berkeley National Laboratory; Email: \texttt{linlin@math.berkeley.edu}} \orcidlink{0000-0001-6860-9566}}

\date{}



\maketitle


\begin{abstract} 
Quantum signal processing (QSP) provides a representation of scalar polynomials of degree $d$ as products of matrices in $\mathrm{SU}(2)$, parameterized by $(d+1)$ real numbers known as phase factors. QSP is the mathematical foundation of quantum singular value transformation (QSVT), which is often regarded as one of the most important quantum algorithms of the past decade, with a wide range of applications in scientific computing, from Hamiltonian simulation to solving linear systems of equations and eigenvalue problems. In this article we survey recent advances in the mathematical and numerical analysis of QSP. In particular, we focus on its generalization beyond polynomials, the computational complexity of algorithms for phase factor evaluation, and the numerical stability of such algorithms. The resolution to some of these problems relies on an unexpected interplay between QSP, nonlinear Fourier analysis on $\mathrm{SU}(2)$, fast polynomial multiplications, and Gaussian elimination for matrices with displacement structure.
\end{abstract}

\section{Introduction.}

Quantum computing has emerged as a new paradigm with the potential to transform many areas of scientific computation. At the most basic level, a quantum computer manipulates information carried by quantum bits (qubits) using quantum gates, each described by a unitary matrix. A quantum algorithm can therefore be viewed as the product of a sequence of unitary transformations together with measurements.  This leads to a simple but fundamental question: how does one design a quantum procedure that evaluates a scalar polynomial, not by adding terms as in the classical setting, but through a product of unitary matrices?

The study of the representation of polynomials has a long history, spanning areas of mathematics such as approximation theory, harmonic analysis, algebraic geometry, and number theory.  It is thus striking that quantum signal processing (QSP) provides a new way to represent polynomials. Originally introduced by Low and Chuang~\cite{LowChuang2017} and subsequently generalized by Gily\'en \textit{et al.}~\cite{GilyenSuLowEtAl2019} and by Haah~\cite{Haah2019}, QSP provides a class of product representations of polynomials that are compatible with the structure of quantum computation, and offers an elegant solution to the question posed above.  
\begin{equation}
X=\begin{pmatrix}
0 & 1\\
1 & 0
\end{pmatrix},
\quad 
Z=\begin{pmatrix}
1 & 0 \\
0 & -1
\end{pmatrix}
\end{equation}
For any choice of \emph{phase factors} $\Psi:=(\psi_0,\psi_1,\cdots,\psi_d)\in\RR^{d+1}$, define
\begin{equation}\label{eqn:qsp-unitary}
    U_d(x, \Psi) := e^{\I \psi_0 Z} \prod_{j=1}^{d} \left[ W(x) e^{\I \psi_j Z} \right], \quad W(x) = e^{\I \arccos(x) X}=\left(\begin{array}{cc}{x} & {\I \sqrt{1-x^{2}}} \\ {\I \sqrt{1-x^{2}}} & {x}\end{array}\right).
\end{equation}
The term ``signal processing'' originated from an analogy with digital filter design on classical computers.  Here $x\in[-1,1]$ and can be viewed as a ``signal'' encoded by the matrix $W(x)$ in $\mathrm{SU}(2)$, which is the group of $2\times 2$ unitary matrices with determinant $1$. These signals are interleaved with a sequence of Pauli $Z$ rotations parameterized by the phase factors $\Psi$. For each $x\in[-1,1]$, $U_d(x, \Psi)$ is a matrix in $\mathrm{SU}(2)$. In quantum computing, $U_d$ can be implemented using a standard quantum gate sequence acting on a single qubit.

For $x \in \RR$, the collection of all real polynomials of finite degree forms the polynomial ring $\RR[x]$, and likewise, the collection of complex polynomials forms the ring $\CC[x]$. A straightforward calculation shows that $[U_d(x,\Psi)]_{1,1}$ (the $(1,1)$ entry of the unitary matrix $U_d(x,\Psi)$) lies in $\CC[x]$.

The essence of the QSP representation is the converse statement: given a polynomial $f(x)$ defined on $[-1,1]$ that satisfies certain structural conditions, there exists a sequence of phase factors $\Psi$ such that, for all $x \in [-1,1]$, the polynomial $f(x)=[U_d(x,\Psi)]_{1,1}$ when $f \in \CC[x]$, or as the real (or imaginary) part of the entry when $f \in \RR[x]$.

In this survey, we discuss the following questions related to QSP:
\begin{enumerate}
  \item Given a polynomial $f(x)$, what conditions are needed for the QSP representation to hold?  If such phase factors exist, are they unique? 
  \item Is QSP a fundamentally new representation of polynomials, or is it related to other areas of mathematics? Can QSP be generalized to represent functions beyond polynomials?
  \item How to compute the phase factors efficiently and accurately? What is the optimal complexity?  Is the algorithm numerically stable? 
  \item How to use QSP to design efficient quantum algorithms for tasks in scientific computation?
\end{enumerate}

The remainder of this article is organized as follows. In \cref{sec:examples}, we present illustrative examples of polynomials that admit a QSP representation. \Cref{sec:existence} through \cref{sec:iqsp} address the first two questions above. We highlight that nonlinear Fourier analysis on $\mathrm{SU}(2)$ provides a natural framework particularly for understanding the second question. In \cref{sec:Weiss_algorithm}, we introduce the Weiss algorithm, which can be viewed as a ``matrix completion'' procedure for constructing $U_d$ from partial information about $f$. 

\Cref{sec:algorithms} and \cref{sec:numerical_stability} address the third question, i.e., algorithms for computing phase factors from $U_d$. The inverse nonlinear Fourier transform (inverse NLFT) provides a unified perspective, and several new algorithms have been developed recently based on this connection.  In particular, the inverse nonlinear fast Fourier transform (inverse nonlinear FFT) algorithm achieves the \emph{near optimal} time complexity of $\Or(d \log^2 d)$.
The numerical stability of these algorithms, however, is a highly non-trivial issue. We show that their numerical stability is connected to both the choices in the matrix completion process above, and the stability of Gaussian elimination on matrices with displacement structure. In \cref{sec:iterative_algorithms}, we present iterative algorithms that provide a complementary perspective for finding phase factors.  For the fourth question, \cref{sec:qsvt} briefly reviews quantum singular value transformation (QSVT) and its applications. Finally, \cref{sec:generalizations_outlook} discusses several generalizations of QSP and provides an outlook on future directions.

\section{Examples of QSP representations.}\label{sec:examples}

In this section, we present a few examples of polynomials that can be represented using QSP. These examples can be reproduced using the
\href{https://github.com/qsppack/QSPPACK}{QSPPACK} package \footnote{QSPPACK is an open-source software package for computing phase factors. It is implemented both in MATLAB \url{https://github.com/qsppack/QSPPACK} and in Python \url{https://qsppack.readthedocs.io}.}.

\paragraph{Chebyshev polynomials:} The Chebyshev polynomials of the first kind, denoted by $T_d(x)$, are defined by the relation $T_d(\cos \theta) = \cos(d\theta)$. By choosing $\Psi=(0,0,\cdots,0)\in\RR^{d+1}$, we have
\begin{equation}
U_d(x,\Psi) = e^{\I d \theta X}=
\begin{pmatrix}
{\cos(d\theta)} & {\I \sin(d\theta)} \\ {\I \sin(d\theta)} & {\cos(d\theta)}
\end{pmatrix}, \quad x=\cos \theta.
\end{equation}
Thus $T_d(x)=\Re[U_d(x,\Psi)]_{1,1}$.

\paragraph{All-zero function:} The constant zero function can be represented using QSP with an arbitrarily long sequence of phase factors. From
\begin{equation}
e^{\I \frac{\pi}{4} Z} e^{\I d \theta X} e^{\I \frac{\pi}{4} Z}=
\begin{pmatrix}
{\I \cos(d\theta)} & {\I \sin(d\theta)} \\ {\I \sin(d\theta)} & {-\I \cos(d\theta)}
\end{pmatrix},
\end{equation}
we find that $\Re[U_d(x,\Psi)]_{1,1}=0$. So the phase factors can be chosen as $\Psi=(\frac{\pi}{4},0,\cdots,0,\frac{\pi}{4})$.

\begin{figure}[H]
\begin{center}
    \includegraphics[width=\textwidth]{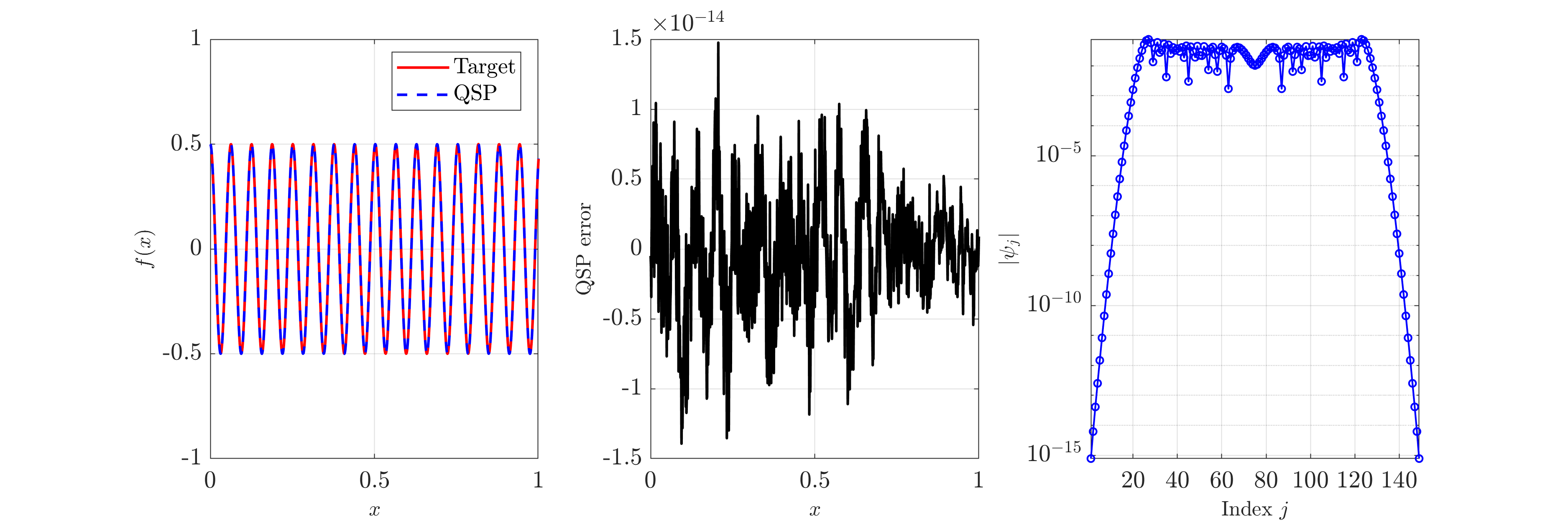}
\end{center}
\caption{QSP representation of $f(x)=\frac12 \cos(100 x)$ using an even polynomial $p(x)$ of degree $150$. Left: the target function and the QSP representation of $p(x)$. Middle: Error between $p(x)$ and its QSP representation. Right: phase factors after removing a factor of $\pi/4$ on both ends plotted on a log scale.}
\label{fig:qsp_cos100x}
\end{figure}

\paragraph{Trigonometric functions:} Consider 
$f(x)=\frac12 \cos(100 x)$. We can first approximate $f(x)$ by an even polynomial $p(x)$ using Chebyshev interpolation, and then use the fixed point iteration (FPI) algorithm in \cref{sec:iterative_algorithms} to find phase factors $\Psi$ such that $\Re[U_d(x,\Psi)]_{1,1}=p(x)$. 
\cref{fig:qsp_cos100x} shows one such polynomial with degree $150$. The QSP error (defined as the difference between the QSP representation and $p(x)$) is about $1.5\times 10^{-14}$, which is close to machine precision. 
After removing a factor of $\pi/4$ on both ends, the phase factors are symmetric with respect to the center of the interval and decay rapidly away from the center. Such a decay behavior can be explained using the infinite quantum signal processing (iQSP) framework discussed in \cref{sec:iqsp}. This example is related to the Hamiltonian simulation problem in \cref{sec:qsvt}.

\begin{figure}[H]
\begin{center}
    \includegraphics[width=\textwidth]{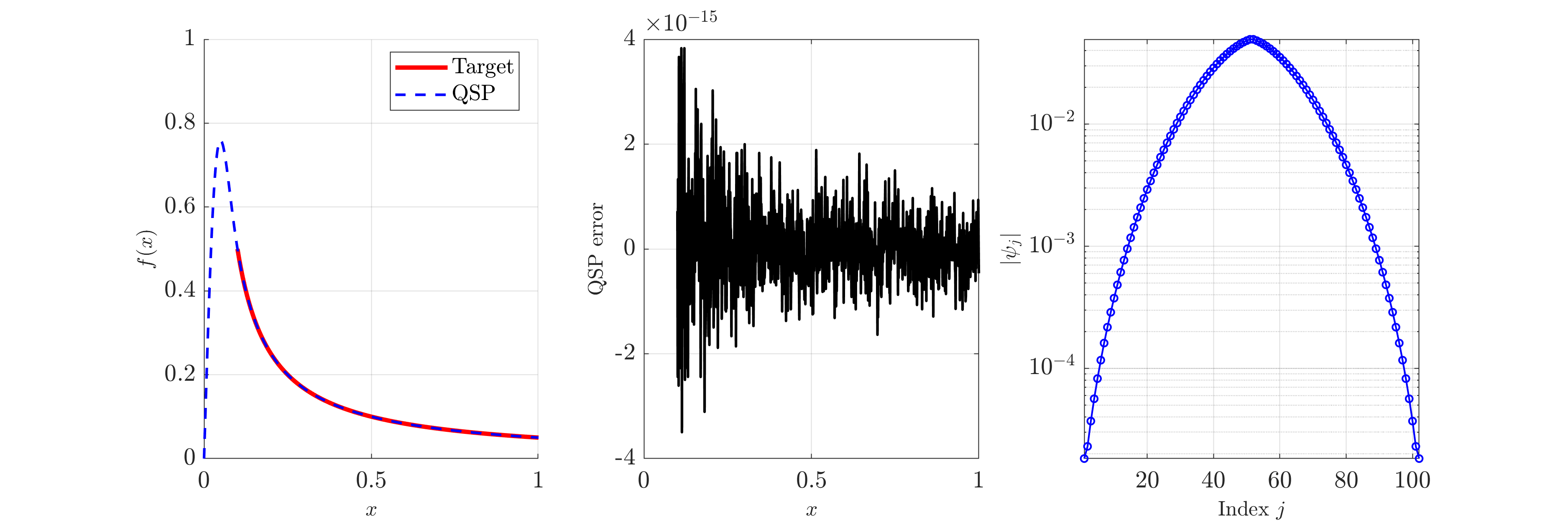}
\end{center}
\caption{Approximating $f(x)=\frac{1}{2\kappa x}$ on $[\kappa^{-1},1]$ using an odd polynomial $p(x)$ of degree $101$ and its QSP representation. Left: the target function and the QSP representation of $p(x)$. Middle: Error between $p(x)$ and its QSP representation. Right: phase factors after removing a factor of $\pi/4$ on both ends plotted on a log scale.}
\label{fig:qsp_inv}
\end{figure}

\paragraph{Inverse function:} We would like to approximate the inverse function $f(x) = \frac{1}{2\kappa x}$ by an odd polynomial $p(x)$ on the interval $[\kappa^{-1}, 1]$, and represent this polynomial using QSP. For $\kappa=10$, 
\cref{fig:qsp_inv} shows one such odd polynomial of degree $d=101$ that is bounded by $1$ on $[-1,1]$. The QSP error is close to machine precision. The phase factors are symmetric with respect to the center of the interval after removing a factor of $\pi/4$ on both ends and decay rapidly away from the center. This example is related to the quantum linear system problem in \cref{sec:qsvt}.

\section{Existence and uniqueness of phase factors.}\label{sec:existence}

For a real or complex function $f$ over $[-1,1]$, if $f$ can be represented by the $(1,1)$ entry of a unitary matrix,  it is necessary that 
\begin{equation}
\norm{f}_\infty := \esssup\limits_{x\in[-1,1]}  \abs{f(x)} \le 1.
\label{eqn:f_infty_bound}
\end{equation}

\begin{theorem}[Quantum signal processing {\cite[Theorem 4]{GilyenSuLowEtAl2019}}]\label{thm:qsp}
    For any $P, Q \in \mathbb{C}[x]$, positive integer $d$ such that
    \begin{enumerate}[label=(\arabic*)]
    \item $\deg(P) \leq d, \deg(Q) \leq d-1$,
    \item $P$ has parity $(d\mod2)$ and $Q$ has parity $(d-1 \mod 2)$,
    \item $|P(x)|^2 + (1-x^2) |Q(x)|^2 = 1, \quad \forall x \in [-1, 1]$,
\end{enumerate}
there exists a set of phase factors $\Psi := (\psi_0, \cdots, \psi_d) \in [-\pi, \pi)^{d+1}$ such that
\begin{equation}
\label{eq:qsp-gslw}
        U_d(x, \Psi) = e^{\I \psi_0 Z} \prod_{j=1}^{d} \left[ W(x) e^{\I \psi_j Z} \right] = \left( \begin{array}{cc}
        P(x) & \I Q(x) \sqrt{1 - x^2}\\
        \I \overline{Q}(x) \sqrt{1 - x^2} & \overline{P}(x)
        \end{array} \right).
\end{equation}
\end{theorem}
Here, $\overline{P}$ is the complex conjugate of $P$. In many applications, we are only interested in using the real part of $P$. Direct calculation shows that for any set of phase factors $\Psi\in [-\pi, \pi)^{d+1}$, the conditions (1)-(3) in \cref{thm:qsp} are satisfied. Therefore, the conditions (1)-(3) are both necessary and sufficient for the existence of phase factors $\Psi$ such that \cref{eq:qsp-gslw} holds. In particular, the condition (3) is simply a normalization condition derived from the unitarity of $U_d(x,\Psi)$.

\begin{corollary}[Quantum signal processing with real target polynomials {\cite[Corollary 5]{GilyenSuLowEtAl2019}}]\label{cor:complementary}
Let $f\in \RR[x]$ be a degree-d polynomial for some $d\geq 1$ such that 
\begin{enumerate}[label=(\arabic*)]
    \item $f(x)$ has parity $(d \mod 2)$,
    \item $\norm{f}_\infty\le 1$. 
\end{enumerate}
Then there exists some $P,Q\in \CC[x]$ satisfying properties (1)-(3) in \cref{thm:qsp} such that $f(x)=\Re[P(x)]$.
\end{corollary}

Note that
\begin{equation}\label{eq:re_im_equiv}
    \Re[U_d(x,\Psi)]_{1,1}=\Im[e^{\I \frac{\pi}{4}Z}U_d(x,\Psi)e^{\I \frac{\pi}{4}Z}]_{1,1}.
\end{equation}
Thus the real part of $[U_d(x,\Psi)]_{1,1}$ can be recovered from the imaginary part by adding $\frac{\pi}{4}$ to both $\psi_0$ and $\psi_d$, and the conclusion of \cref{cor:complementary} also holds if we replace $\Re[P(x)]$ by $\Im[P(x)]$.

Due to the parity constraint, the number of degrees of freedom in a given target polynomial $f\in\RR[x]$ is only $\wt{d}:=\lceil \frac{d+1}{2} \rceil$. Therefore the phase factors $\Psi$ cannot be uniquely defined.
Since we are interested in the top-left entry of $U$, i.e., the polynomial $P$, without loss of generality we may restrict $Q\in\RR[x]$. In such a case, the phase factors can be restricted to be \emph{symmetric}: $\Psi=(\psi_0,\psi_1,\cdots,\psi_1,\psi_0)$.
Let $D_d$ denote the domain of the symmetric phase factors:
\begin{equation}
    D_d=\begin{cases}
    [-\frac{\pi}{2},\frac{\pi}{2})^{\frac{d}{2}} \times [-\pi,\pi) \times [-\frac{\pi}{2},\frac{\pi}{2})^{\frac{d}{2}}, & d \mbox{ is even,}\\
    [-\frac{\pi}{2},\frac{\pi}{2})^{d+1}, &d \mbox{ is odd.}\\
    \end{cases}
\end{equation}
The number of degrees of freedom in $D_d$ is exactly $\wt{d}$, which matches the number of degrees of freedom in $f$. These effective degrees of freedom are called the \emph{reduced phase factors}.

\begin{theorem}[Quantum signal processing with symmetric phase factors {\cite[Theorem 1]{WangDongLin2022}}]
\label{thm:sym_qsp}
For any $P, Q \in \mathbb{C}[x]$, positive integer $d$ such that
\begin{enumerate}[label=(\arabic*)]
    \item $\deg(P)= d$ and $\deg(Q)= d-1$.
    \item $P$ has parity $(d \bmod 2)$ and $Q$ has parity $(d-1 \bmod 2)$.
    \item $|P(x)|^2 + (1-x^2) |Q(x)|^2 = 1, \forall x \in [-1, 1]$,
    \item \label{itm:4} If $d$ is odd, then the leading coefficient of $Q$ is positive,
\end{enumerate}
there exists a unique set of symmetric phase factors $\Psi:=(\psi_0,\psi_1,\cdots,\psi_1,\psi_0)\in D_d$ such that 
\begin{equation}\label{eq:UPQ}
U_d(x,\Psi)=\begin{pmatrix}
P(x) & \I Q(x)\sqrt{1-x^2}\\
\I Q(x) \sqrt{1-x^2} & \overline{P} (x)
\end{pmatrix}.
\end{equation}
\end{theorem}
When we are only interested in $f(x)=\Re[P(x)]$ or $f(x)=\Im[P(x)]$ represented by symmetric phase factors, the conditions on $f$ are the same as those in \cref{cor:complementary}. 

We emphasize that the set of symmetric phase factors is unique \emph{only if} both $P(x)$ and $Q(x)$ are determined.  If only $f(x)=\Re[P(x)]$ or $f(x)=\Im[P(x)]$ is given, then the set of symmetric phase factors is generally not unique. In fact, the number of solutions grows \emph{combinatorially} as $d$ increases~\cite[Theorem 4]{WangDongLin2022}.

Surprisingly, there is one solution that stands out among the combinatorially many solutions, and enjoys many desirable properties. We follow~\cite{WangDongLin2022} and refer to it as the \emph{maximal solution}, and will discuss its properties in the next few sections. For now we just note that all solutions shown in \cref{sec:examples} are maximal solutions.

\section{Connections to nonlinear Fourier analysis on \texorpdfstring{$\mathrm{SU}(2)$}{SU(2)}.}\label{sec:nlft}

Is QSP a fundamentally new representation of polynomials, or is it related to other areas of mathematics? Like many other mathematical discoveries, the idea of representing polynomials using products of matrices has been reinvented multiple times in different contexts, and can be categorized as a special case of \emph{nonlinear Fourier analysis} (we refer readers to \cite{tao2012nonlinear} by Tao and Thiele for an introductory treatment). The Fourier transform is a fundamental tool in mathematics and is used ubiquitously in scientific and engineering computations. In comparison, the nonlinear Fourier transform (NLFT) is far less well-known.
Its origin can be traced back to Schur's 1917 study of the properties of bounded holomorphic functions on the unit disk, now known as Schur functions~\cite{schur1918potenzreihen}. Over the following century, NLFT has been rediscovered in seemingly unrelated contexts under different names, including scattering theory~\cite{beals1985inverse,winebrenner1998linear,damanik2004half,hitrik2001properties}, integrable systems~\cite{tanaka1972some,fokas1994integrability}, orthogonal polynomials~\cite{szego1939ortho,case1975orthogonal,denisov2002probability,deift2000orthogonal}, Jacobi matrices~\cite{simon2004canonical,killip2003sum,damanik2004half,volberg2002inverse}, logarithmic integrals~\cite{koosis1998logarithmic}, and stationary Gaussian processes~\cite{dym2008gaussian}, to name a few.

Briefly speaking, NLFT replaces the addition operation in the linear Fourier transform with matrix multiplication. It maps a complex sequence $\bga=(\gamma_k)_{k\in\ZZ}$ to a one-parameter family of  matrices $\overbrace{\bga}(z)$, where $z$ is on the unit circle $\mathbb{T}$. Moreover, $\overbrace{\bga}(z)$ can be expressed as a product of $z$-dependent matrices, where each matrix is in a Lie group and is parameterized by an entry $\gamma_k$. In the case of Schur functions and the aforementioned applications, this Lie group is 
\begin{equation}
\mathrm{SU}(1,1):=\Set{\begin{pmatrix}
a & b\\
\overline{b} & \overline{a}
\end{pmatrix} : \abs{a}^2-\abs{b}^2=1, \quad a,b\in\CC},
\end{equation}
The transformation from $\bga$ to $\overbrace{\bga}$ is called the forward NLFT, and the mapping from $\overbrace{\bga}$ back to $\bga$ is called the inverse NLFT. 

Compared to the $\mathrm{SU}(1,1)$ case, the NLFT on $\mathrm{SU}(2)$
has been studied much later. This was first systematically explored in the thesis of Tsai \cite{tsai2005nlft}, which derives analytic results in the $\mathrm{SU}(2)$ setting that parallel those of $\mathrm{SU}(1,1)$. However, there are important differences between these two cases, particularly in terms of the domain,  range, and injectivity of the NLFT map (for example, compare~\cite[Theorem~2.3]{tsai2005nlft} and~\cite[Theorem~1]{tao2012nonlinear} for the relevant results for compactly supported sequences). The NLFT on $\mathrm{SU}(2)$ has applications in the study of solitons from certain nonlinear Schr\"{o}dinger equations (see e.g., \cite{faddeev1987hamiltonian},~\cite[Chapter~5]{tsai2005nlft}), but QSP is arguably its most significant application so far. The connection between QSP and the $\mathrm{SU}(2)$ NLFT was recently established by~\cite{alexis2024quantum}, which showed that determining the phase factors in QSP is equivalent to solving a variant of the inverse NLFT.

Throughout the rest of the discussions, the unit circle is denoted by $\TT$, the open unit disk is denoted by $\DD$, and the closed unit disk by $\cDD$. The Riemann sphere is $\CC \cup \{\infty\}$, and we define $\CC^\ast := \CC \setminus \{0\}$. For a Laurent polynomial $a(z)$, define $a^*(z) := \overline{a(\overline{z}^{-1})}$ for $z\in\CC \cup \{\infty\}$.  Let $\bga: \ZZ \rightarrow \CC$ be a compactly supported sequence, and we will denote the space of all such sequences by $\lzero$.  For a $\bga \in \lzero$, whose support lies in $[m,n]$ with $m,n \in \ZZ$, the \emph{nonlinear Fourier transform} of $\bga$ is defined as a finite product of matrix-valued meromorphic functions:
\begin{equation}
\label{eq:nonlinear-Fourier-series}
    \overbrace{\bga}(z) := \prod_{k=m}^{n} \left[\frac{1}{\sqrt{1 + |\gamma_k|^2}} 
    \begin{pmatrix}
        1 & \gamma_k z^k \\
        - \ol{\gamma_k} z^{-k} & 1
    \end{pmatrix}\right], \quad z\in\CC \cup \{\infty\}.
\end{equation}
Taking the determinant of the matrix factors 
$(1 + |\gamma_k|^2)^{-1/2} \left(
\begin{smallmatrix}
        1 & \gamma_k z^k \\
        - \ol{\gamma_k} z^{-k} & 1
\end{smallmatrix} \right)$ 
appearing in \cref{eq:nonlinear-Fourier-series}, we see that the determinant of each factor in the product and also of $\overbrace{\bga}(z)$ is $1$ everywhere on the Riemann sphere, by analytic continuation. Moreover, the matrix factors are elements of $\mathrm{SU}(2)$ when $z \in \TT$, and thus so is $\overbrace{\bga}(z)$.

When  $\sum_{k\in\ZZ}\abs{\gamma_k}$ is small, the NLFT of $\bga$ can be approximated by its linear approximation:
\begin{equation}
\overbrace{\bga}(z) \approx\begin{pmatrix}
    1 & \sum_{k=m}^{n} \gamma_k z^k \\
    -\sum_{k=m}^{n} \overline{\gamma_k} z^{-k} & 1
\end{pmatrix}.
\end{equation}
Therefore, the standard Fourier series can be viewed as the leading order contribution to the upper-right entry of $\overbrace{\bga}(z)$. When the coefficients $\gamma_k$ are not small, the difference between the two quantities can become significant.

Note that the definition $a^*(z) := \overline{a(\overline{z}^{-1})}$  implies $(a^\ast)^\ast = a$, and $(ab)^\ast = a^\ast b^\ast$. For instance, if $a(z)$ is a finite series of the form $a(z) = \sum_{k=0}^{n} \alpha_k z^{\beta_k}$, where $\alpha_k \in \CC$ and $\beta_k \in \ZZ$, for each $k$, then $a^*(z) = \sum_{k=0}^{n} \overline{\alpha_k} \left( \frac{1}{z} \right)^{\beta_k}$. Thus $\overbrace{\bga}(z)$ is always of the form
\begin{equation}
\label{eq:nlft-ab-def}
    \overbrace{\bga}(z) = 
    \begin{pmatrix}
            a(z) & b(z)\\ -b^*(z) & a^*(z)
    \end{pmatrix},
\end{equation}
where $a(z)$, and $b(z)$ are Laurent polynomials. Thus we may omit the second row of the matrix and denote (with a slight abuse of notation) $\overbrace{\bga}:= (a, b)$.

\begin{theorem}[NLFT bijection {\cite[Section~3]{alexis2024quantum}, \cite[Chapter~2]{tsai2005nlft}}]
\label{lem:nlft-bijection}
The NLFT is a bijection from $\lzero$ onto the space
\begin{equation}
\label{eq:nlft-image}
\mathcal{S} = \{(a,b): a,b \text{ are Laurent polynomials}, \; aa^\ast + bb^\ast = 1, \; 0 < a^\ast(0) < \infty \}.
\end{equation}
\end{theorem}
From the bijective property of NLFT, we may define the problem of computing the \textit{inverse nonlinear Fourier transform}, i.e., 
for a given $(a,b) \in \mathcal{S}$, compute the unique $\bga \in \lzero$ such that $\overbrace{\bga} = (a,b)$.

The problem of determining the phase factors in QSP can be viewed as a special case of the inverse NLFT problem. The result below is from \cite[Lemma 3.1]{ni2025inverse}, which is a variant of \cite[Lemma~1]{alexis2024quantum}.

\begin{lemma}[Connection between QSP and NLFT]
\label{lem:HUdH-lemma-nonsym}
For any $d\in\NN$ and $\Psi := (\psi_0, \cdots, \psi_d) \in [-\pi, \pi)^{d+1}$, define the sequence $\bga: \ZZ \rightarrow \CC$ as $\gamma_k := \tan \psi_k$, for $k=0,\dots,d$, and zero otherwise. Then for all $\theta \in [0,\pi]$ we have
    \begin{equation}
    \label{eq:HUdH-lemma-nonsym-1}
    \begin{pmatrix}
        1 & 0 \\
        0 & i
    \end{pmatrix}
    \mathrm{H}\; U_d(\cos \theta,\Psi)\; \mathrm{H} 
    \begin{pmatrix}
        1 & 0 \\
        0 & -i
    \end{pmatrix}
    = \overbrace{\bga}( e^{2i\theta}) 
    \begin{pmatrix}
        e^{i d \theta} & 0 \\
        0 & e^{-i d \theta}
    \end{pmatrix}, \quad 
    \mathrm{H}:= \frac{1}{\sqrt{2}}  
    \begin{pmatrix}
    1 & 1 \\1 & -1
    \end{pmatrix}.
    \end{equation}
    where $U_d(x,\Psi)$ is defined in \cref{eqn:qsp-unitary} with $x=\cos\theta$, and $\mathrm{H}$ is the Hadamard matrix.
\end{lemma}

Define a set $\mathcal{B}$ as the projection of $\mathcal{S}$ onto the second component:
\begin{equation}
\mathcal{B} = \{b ~|~  \exists\, \text{ Laurent polynomial } a \text{ such that } (a,b)\in\mathcal{S} \,\}.
\end{equation}
If a degree $d$ real polynomial $f$ can be expressed as
\begin{equation}
\label{eq:b_f_relation}
    f(\cos\theta) = \Re[b(e^{2i \theta}) e^{-i d \theta} ],   \quad \forall \theta\in[0,\pi],
\end{equation}
for some $b\in \mathcal{B}$, then the phase factors $\Psi$ can be determined by solving the inverse NLFT problem. In particular, if $\overbrace{\bga}=(a,b)\in\mathcal{S}$, then the phase factors $\psi_k = \arctan\gamma_k$ for $k=0,\dots,d$ satisfy (see~\cite[Theorem 3.2]{ni2025inverse})
\begin{equation}
    f(x) = \Im[U_d(x,\Psi)]_{1,1}, \quad \forall x\in[-1,1].
\end{equation}

\section{Infinite quantum signal processing and convergence properties.}\label{sec:iqsp}

The problem of \emph{infinite quantum signal processing} (iQSP) asks whether the QSP representation can be extended to non-polynomial functions $f$ through a product of \emph{countably many} unitary matrices.

For simplicity, we assume that the target function $f$ is a real-valued measurable and even function satisfying
\cref{eqn:f_infty_bound}.  One natural idea is to consider the limit of a sequence of polynomials $\{f^{(2d)}\}_{d=1}^\infty$ that converges to $f$ as $d\to\infty$ in some sense, and for each $f^{(2d)}$ we find a set of phase factors $\Psi^{(2d)}$. One immediate issue is that the phase factors $\Psi^{(2d)}$ are not unique, and it is not clear how to choose a sequence of phase factors $\{\Psi^{(2d)}\}_{d=1}^\infty$ such that the limit $\lim_{d\to\infty}\Psi^{(2d)}$ exists.

Let $\mathbf{P}$ denote the space of infinite sequences
$\Phi=(\psi_k)_{k\in \NN}$ with $\psi_k\in [-\pi/2,\pi/2]$. 
We equip $\mathbf{P}$ with a metric induced by the $\ell^\infty$ norm
$\norm{\Phi}_{\infty} := \sup_{k\in \NN} \abs{\psi_k}$.

Given any $\Phi \in \mathbf{P}$ and $x\in[0,1]$, one can define a sequence of unitary matrices using the following recursive relation:
\begin{equation}
\begin{split}
    &V_0(x,\Phi) = e^{\I \psi_0 Z}
    \\
    &V_d(x,\Phi) = e^{\I \psi_{d} Z}W(x) V_{d-1}(x,\Phi)W(x) e^{\I \psi_d Z}.
\end{split}
\label{eqn:sym_qsp}
\end{equation}
It is easy to see that this corresponds to symmetric phase factors $\Psi^{(2d)}$ of the form $\Psi^{(2d)}=(\psi_d,\psi_{d-1},\ldots,\psi_1,\psi_0,\psi_1,\ldots,\psi_{d-1},\psi_{d})\in \RR^{2d+1}$, such that
$V_d(x,\Phi)=U_{2d}(x,\Psi^{(2d)})$. So $\Phi$ can be viewed as the reduced phase factors in the infinite dimensional case.

Let $u_d(x,\Phi)=V_d(x,\Phi)_{1,1}$.  In order to compare phase factors of different lengths, an important observation is that for $\Phi=(\psi_0,\psi_1,\ldots,\psi_{d},0,0,\ldots)$, for any $n\ge d$, $\Im[u_n(x,\Phi)]=\Im[u_{d}(x,\Phi)]$~\cite[Lemma 10]{DongLinNiEtAl2024_iqsp}. Therefore, a polynomial $f$ can be meaningfully represented by an infinite sequence $\Phi\in\mathbf{P}$ such that $f(x)=\lim_{d\to\infty} \Im[u_d(x,\Phi)]$ for all $x\in[-1,1]$.

\subsection{\texorpdfstring{$L^1$}{L1} convergence.}

Ref.~\cite{DongLinNiEtAl2024_iqsp} provided the first infinite QSP construction.
\begin{theorem}[Infinite QSP {\cite[Theorem 3]{DongLinNiEtAl2024_iqsp}}]
\label{thm:iqsp_l1_conv}
For any real, even function $f$ satisfying \cref{eqn:f_infty_bound} with the Chebyshev expansion $f(x)=\sum_{k=0}^{\infty} c_k T_{2k}(x)$, if the $\ell^1$ norm of the Chebyshev coefficient $\norm{\vc}_{1}:=\sum_{k} \abs{c_k}\le 0.9$, then there exists a sequence $\Phi\in\ell^1(\NN)\subset \mathbf{P}$ such that
\begin{equation}
\lim_{d\to \infty}  \sup_{x\in[-1,1]}\abs{\Im[u_d(x,\Phi)] - f(x)} = 0.
\label{eqn:infnorm_conv}
\end{equation}
\end{theorem}

Eq.~\eqref{eqn:infnorm_conv} immediately implies convergence in $L^1$ norm.
\begin{equation}
\lim_{d\to \infty} \int_{-1}^1 \abs{\Im[u_d(x,\Phi)] - f(x)} \ud x = 0.
\label{eqn:l1_conv}
\end{equation}
Under the same $\ell^1$ condition on the Chebyshev coefficients, we can obtain a Lipschitz bound, which establishes the stability of the inverse map from $f$ to $\Phi$.
\begin{theorem}[{\cite[Theorem 24]{DongLinNiEtAl2024_iqsp}}]
\label{thm:iqsp_lip}
For two functions $f(x)=\sum_{k=0}^{\infty} c_k T_{2k}(x), f'(x)=\sum_{k=0}^{\infty} c'_k T_{2k}(x)$ with $\norm{\vc}_1,\norm{\vc'}_1\le 0.9$, let $\Phi, \Phi'\in\ell^1(\NN)$ be the corresponding sequences in \cref{thm:iqsp_l1_conv}, then we have the Lipschitz bound
\begin{equation}
C_1 \norm{\vc-\vc'}_1\le \norm{\Phi-\Phi'}_1 \le C_2 \norm{\vc-\vc'}_1,
\end{equation}
for some constants $C_1,C_2>0$. 
\end{theorem}

In particular, for $f(x)=\sum_{k=0}^{\infty} c_k T_{2k}(x)$, let $\Phi=(\psi_0,\psi_1,\ldots)\in\ell^1(\NN)$ be the corresponding sequence in \cref{thm:iqsp_l1_conv}. Then for any $n\ge 0$, setting $\Phi'=(\psi_0,\ldots,\psi_n,0,\ldots)$, we have the following decay estimate on the tail of $\Phi$~\cite[Theorem 4]{DongLinNiEtAl2024_iqsp}:
\begin{equation}\label{eq:ineq-weak-decay}
    C_1\sum_{k>n} \abs{c_k} \le \sum_{k>n}\abs{\psi_k}\le C_2\sum_{k>n} \abs{c_k}.
\end{equation}
If the Chebyshev coefficients $c_k$ decay rapidly, then the phase factors $\psi_k$ also decay rapidly with the same rate. This behavior is also observed numerically in \cref{sec:examples}.

The proof of \cref{thm:iqsp_l1_conv} relies on the inverse mapping theorem, and it remains open whether $L^1$ convergence can be established to all functions satisfying \cref{eqn:f_infty_bound}. 

\subsection{\texorpdfstring{$L^2$}{L2} convergence.}\label{sec:L2_conv_iqsp}

Is it possible to generalize these results to a larger class of functions? Alexis, Mnatsakanyan and Thiele~\cite{alexis2024quantum} provided the first answer to this question using nonlinear Fourier analysis. The convention for the integral on $\TT$ is
\begin{equation}
\int_{\TT} g :=\frac{1}{2\pi}\int_{0}^{2\pi}g(e^{i\theta}) \ud\theta.
\end{equation}
If a real-valued measurable even function $f:[0,1]\to \RR$ can be expressed as 
\begin{equation}
    f(\cos\theta)=g(e^{i\theta}),  \quad \forall \theta\in[0,2\pi)
\end{equation}
for some function $g$ defined on $\TT$, then
we have
\begin{equation}\label{hsnorm}
\int_{\TT} |g|^2 :=\frac{1}{2\pi}\int_{0}^{2\pi} \abs{f(\cos\theta)}^2 \ud\theta=\frac{2}{\pi}\int_{0}^1 \abs{f(x)}^2 \frac{\ud x}{\sqrt{1-x^2}}:=\left \| f \right \|_{\mathbf{S}}^2.
\end{equation}
Here $\left \| f \right \|_{\mathbf{S}}$ is called the \emph{Szeg\H o norm}. 
A real-valued measurable even function $f:[0,1]\to [-1,1]$ is called a \emph{Szeg\H o function} if it satisfies the following Szeg\H o-type condition:
\begin{equation}\label{eq:Szego}
\int_{0}^1 \log|1-f(x)^2| \frac{\ud x}{\sqrt{1-x^2}} > - \infty \, .
\end{equation}
We use  $\mathbf{S}$ to denote the set of all Szeg\H o functions.
Since $y\le -\log(1-y)$ for all $y\in[0,1)$, the Szeg\H o condition in \cref{eq:Szego} implies that $\norm{f}_{\mathbf{S}}<\infty$.
Note that $\norm{\cdot}_{\mathbf{S}}$ induces an inner product, and $\mathbf{S}$ is a subset of a Hilbert space. In particular, for $f(x)=\sum_{k\in \NN} c_k T_{2k}(x)$, we have $\left \| f \right \|_{\mathbf{S}}^2=\abs{c_0}^2+\frac12 \sum_{k>0} \abs{c_k}^2$. So $\left \| f \right \|_{\mathbf{S}}<\infty$ is equivalent to the square summable condition $\norm{\vc}_{2}:=\sqrt{\sum_{k} \abs{c_k}^2}<\infty$. 

In standard $L^2$ theory of Fourier analysis, the Plancherel identity plays a fundamental role, namely for $f(x)=\sum_{k=0}^{\infty} c_k T_k(x)$, we have
\begin{equation}
\int_{-1}^1 \abs{f(x)}^2 \frac{\ud x}{\sqrt{1-x^2}} = \pi \abs{c_0}^2+\frac{\pi}{2}\sum_{k=1}^{\infty} \abs{c_k}^2.
\end{equation}
A nonlinear analogue of the Plancherel identity on $\mathrm{SU}(2)$ is established in \cite[Theorem 1]{alexis2024quantum}.

\begin{theorem}[Infinite QSP, $L^2$ convergence {\cite[Theorem 1]{alexis2024quantum}}]\label{thm:iqsp_L2_partial}
For each $f \in \mathbf{S}$ and $0<\eta<\frac{1}{\sqrt{2}}$ satisfying
\begin{equation}\label{eq:eta_condition_partial}
\norm{f}_{\infty} \le \frac{1}{\sqrt{2}} - \eta,
\end{equation}
there exists a unique sequence $\Phi=(\psi_k)_{k\in\NN} \in \mathbf{P}$ such that 
\begin{equation}
\lim_{d\to \infty} \norm{\Im[u_d(x,\Psi)]-f(x)}_{\mathbf{S}}=0,
\label{eqn:L2_conv}
\end{equation}
and the following nonlinear Plancherel identity holds:
\begin{equation}\label{eqn:plancherel_nonlinear}
-\frac{2}{\pi}\int_{0}^1 \log|1-f(x)^2| \frac{\ud x}{\sqrt{1-x^2}}=\sum_{k\in \ZZ}\log(1+\tan^2\psi_{|k|}).
\end{equation}
Furthermore, for two functions $f, f' \in \mathbf{S}$ satisfying \cref{eq:eta_condition_partial} with corresponding sequences $\Phi, \Phi'$ as above, we have the Lipschitz bound
\begin{equation}\label{eq:Lip_bd_QSP}
\left \| \Phi - \Phi ' \right \|_{\infty} \leq 7.3 \eta^{-\frac{3}{2}} \left \| f - f' \right \|_{\mathbf{S}}.
\end{equation}
\end{theorem}

To establish this result, the concept of NLFT needs to be generalized from compactly supported sequences in $\lzero$ to square summable sequences $\bga\in\ell^2(\ZZ)$. First, NLFT can be extended directly from $\lzero$ to square-integrable sequences supported on the half-line $\ell^2(\NN)=:\ell^2 ([0, \infty))$ 
 \cite{alexis2024quantum,tsai2005nlft}. In the latter case, $(a(z), b(z))$ may no longer be a pair of Laurent polynomials. 
For $k\in \ZZ$, the image  of $\ell^2 ([k, \infty))$ under the NLFT is denoted by $\mathbf{H}_{\geq k}$ and is characterized in \cite{tsai2005nlft} and \cite[Section 6]{alexis2024quantum}. 

The extension to sequences in $\ell^2 (\ZZ)$ as follows: given a sequence $\bga$ in $\ell^2(\ZZ)$, split it as $\bga_-+\bga_+$, where $\bga_-$ is supported in $(-\infty, -1] $  and $\bga_+$ is supported in $[0,\infty)$. Then we define the nonlinear Fourier transform of $\bga$ to be the pair
\begin{equation}\label{eq:factorization_explain}
\overbrace{\bga}=(a,b):= (a_- , b_{-}) (a_+, b_+)
\end{equation}
where $\overbrace{\bga_{-}}=(a_- , b_{-})$ and $\overbrace{\bga_{+}}=(a_+, b_+)$.
The problem of finding factors $(a_{-}, b_{-})$ and $(a_+, b_+)$ as in \cref{eq:factorization_explain} is known as a \emph{Riemann--Hilbert factorization problem}~\cite[Lecture 3, p.31]{tao2012nonlinear}.  The Riemann--Hilbert 
factorization also provides a powerful algorithm for computing phase factors (see \cref{sec:RH_algorithm}).

The proof of \cref{thm:iqsp_L2_partial} relies on solving the Riemann--Hilbert factorization problem by designing a Banach contraction mapping,
and the condition in \cref{eq:eta_condition_partial} is a technical condition used to ensure the contraction property.  We also note that \cite[Theorem 5]{alexis2024quantum} provides another perspective on the $L^1$ convergence of iQSP.

Ref.~\cite{alexis2024infinite} provided a solution of the Riemann--Hilbert factorization problem without relying on the Banach contraction mapping, and established the $L^2$ convergence for all Szeg\H o functions.

\begin{theorem}[Infinite QSP, $L^2$ convergence for all Szeg\H o functions {\cite[Theorem 1]{alexis2024infinite}}]\label{thm:iqsp_L2_full}
For each $f \in \mathbf{S}$ and $0<\eta<\frac{1}{2}$ satisfying
\begin{equation}\label{eq:eta_condition_full}
\norm{f}_{\infty} \le 1 - \eta,
\end{equation}
there exists a unique sequence $\Phi=(\psi_k)_{k\in\NN} \in \mathbf{P}$ such that $\Im[u_d(x,\Psi)]$ converges to $f$ in the sense of \cref{eqn:L2_conv} and $\Phi$ satisfies the Plancherel identity in \cref{eqn:plancherel_nonlinear}. 

Furthermore, for two functions $f, f' \in \mathbf{S}$ satisfying \cref{eq:eta_condition_full} with corresponding sequences $\Phi, \Phi'$ as above, we have the Lipschitz bound
\begin{equation}\label{eq:Lip_bd_QSP_full}
\left \| \Phi - \Phi ' \right \|_{\infty} \leq 1.6 \eta^{-3} \left \| f - f' \right \|_{\mathbf{S}}.
\end{equation}
\end{theorem}

\section{Complementary polynomials and Weiss algorithm.}\label{sec:Weiss_algorithm}

Recall that for any real polynomial $f$ satisfying the conditions in \cref{cor:complementary}, there exists a pair of polynomials $(P,Q)$ such that $f(x)=\Re[P(x)]$. When the phase factors are symmetric, $Q$ is a real polynomial.  The real polynomials $(\Im P,Q)$ are called  the \emph{complementary polynomials} associated with $f$. Due to the connection between QSP and NLFT, the existence of complementary polynomials is equivalent to the problem of finding a Laurent polynomial $a$ such that $(a,b)\in\mathcal{S}$ for $b$ satisfying \cref{eq:b_f_relation}.

On the open unit disk $\mathbb{D}$, a function $g(z)$ is in the Hardy space $H^p(\mathbb{D})$ for $1\le p <\infty$ if $g(z)$ is holomorphic on $\mathbb{D}$ and
\begin{equation}
\sup_{0 \leq r < 1} \int_{0}^{2\pi} |g(re^{i\theta})|^p \ud \theta < \infty \, .
\end{equation}
Similarly, $g \in H^{\infty} (\mathbb{D})$ if 
\begin{equation}
\sup_{0 \leq r < 1}  \sup\limits_{\theta} |g(re^{i\theta})| < \infty.
\end{equation}
Functions in $H^p(\mathbb{D})$ have radial limits almost everywhere on the unit circle $\TT$, and these boundary values determine the function uniquely. Thus when we say a function $g \in L^p (\TT)$ belongs to $H^p (\DD)$, we mean $g$ coincides with the boundary values of a unique $H^p (\DD)$ function almost everywhere, which we also denote by $g$. By the mean value property for harmonic functions, for every function $g \in H^p (\DD)$ we have $g(0) = \int_{\TT} g$.

If $g$ is a periodic smooth function on $\mathbb{T}$, define the \emph{Hilbert transform}, where $\operatorname{p.v.}$ stands for the Cauchy principal value,
\begin{equation}
\mathrm{H}(g)(x) := \frac{1}{\pi} \operatorname{p.v.} \int_{0}^{2\pi} g(e^{i\theta}) \frac12 \cot\left(\frac{x - \theta}{2}\right) \ud \theta.
\end{equation}
Direct calculation shows that
\begin{equation}\label{eq:Hilbert_tranform_rule}
\mathrm{H}(z^n)=-\I z^n, \quad n\in \mathbb{N}_+, \quad
\mathrm{H}(z^{-n})=\I z^{-n}, \quad n\in \mathbb{N}_+.
\end{equation}

A function $g \in L^{\infty} (\TT)$ is called an \emph{outer function}, if 
\begin{equation}
    \log |g| \in L^1 (\TT) \quad \text{and} \quad g=e^{G} \text{ where } G = \log |g| + i \mathrm{H}(\log |g|).
\end{equation}
An outer function $g$ can be analytically continued to $\DD$ with $g \in H^{\infty} (\DD)$. The concept of outer function is important in the construction of numerically stable algorithms. If $g$ is a polynomial, then $g$ is outer if and only if all its roots are outside the unit disk $\DD$.

One immediate reason for introducing the outer function in the present context is that, besides the Plancherel identity in \cref{eqn:plancherel_nonlinear}, there is also a nonlinear Plancherel inequality for NLFT:
\begin{lemma}[Nonlinear Plancherel inequality {\cite[Lemma 15]{alexis2024infinite}}]\label{lem:plancherel_ineq}
If $\overbrace{\bga}=(a,b)$ for some $\bga \in\ell^2 (\ZZ)$, then
\begin{equation}\label{eq:Plancherel_ineq}
- \int_{\TT} \log (1- |b(z)|^2)
\le \sum_{k\in \ZZ}\log(1+|\gamma_k|^2).
\end{equation}
The equality holds if and only if $a^*$ is outer.
\end{lemma}

\cref{thm:construct_a} further states that the choice of an outer function $a^*$ is always possible and is unique, if $b$ satisfies the Szeg\H o condition in \cref{eq:Szego_cond_b}.
This is the reason for being able to establish the uniqueness statement in \cref{thm:iqsp_L2_full}. This choice also coincides with the choice of the \emph{maximal solution} from \cite{WangDongLin2022}. See the discussions in \cite[Section 4.4]{alexis2024infinite}.

\begin{theorem}[{\cite[Theorem 4]{alexis2024infinite}}]\label{thm:construct_a}
For each complex valued measurable function \( b \) on \( \mathbb{T} \) with $\norm{b}_{\infty} \leq 1$,
if $b$ satisfies the Szeg\H o condition 
\begin{equation}\label{eq:Szego_cond_b}
\int\limits_{\TT} \log (1 -|b(z)|^2) > -\infty \, ,
\end{equation}
then there is a unique measurable function \( a \) on \( \mathbb{T} \) such that $a^*$ is outer, $a^*(0) > 0$, and \begin{equation}\label{eq:det_torus_gen} 
aa^*+bb^*=1
\end{equation}
almost everywhere on $\TT$.
\end{theorem}

The proof of \cref{thm:construct_a} is constructive,
which gives rise to the \emph{Weiss algorithm} for constructing $a$ from $b$.
Below we discuss the Weiss algorithm for constructing complementary polynomials as presented in~\cite[Section 2.3]{alexis2024infinite}. The idea and hence the name of the algorithm was derived from the Guido and Mary Weiss algorithm \cite{weiss1962derivation}.

The Weiss algorithm consists of three steps. (1) Compute  $R(z):=\log\sqrt{1-\abs{b(z)}^2}$ (2) Using the Hilbert transform, compute $G(z):=R(z)-\I\mathrm{H}(R(z))$. (3) Evaluate $a(z):= e^{G(z)}$, which satisfies the conditions in \cref{thm:construct_a}. All computations can be done on the unit circle $\TT$, and the Hilbert transform can be efficiently evaluated using the fast Fourier transform (FFT). 

Specifically, we evaluate the degree $d$ Laurent polynomial $b(z)$ on $N$ equally spaced points on $\TT$. If $\sup\limits_{z \in \TT} |b(z)|=1-\eta$ for some $\eta>0$, then we should choose
$N=\mathcal{O}(\frac{d}{\eta} \log(\frac{d}{\eta\epsilon}))$~\cite[Theorem 8]{alexis2024infinite}.  The FFT and its inverse can be used to compute the Hilbert transform in $\Or(N\log N)$ operations. The overall computational cost of the Weiss algorithm is thus
$\Or\left(\frac{d}{\eta}\log^2(d/(\eta\epsilon))\right)$.  We refer to \cite[Section 3.2]{alexis2024infinite} for details of the Weiss algorithm.

The first constructive solution to the problem of finding complementary polynomials was presented in Refs.~\cite{GilyenSuLowEtAl2019,Haah2019}, and later extended in \cite{WangDongLin2022}. This approach involves finding all roots of the Laurent polynomial $1-f((z+z^{-1})/2)^2$. However, the root-finding algorithm requires $\Or(d\log(d/\epsilon))$ bits of precision~\cite{Haah2019}, which makes it \emph{numerically unstable}. Ying developed the first algorithm to directly construct complementary polynomials without root-finding~\cite{Ying2022} using contour integrals. There is another contour integral based approach in \cite{berntson2025complementary}, which is equivalent to the Weiss algorithm. 

\section{Algorithms for inverse nonlinear Fourier transform.}\label{sec:algorithms}

Now that the Laurent polynomial $a$ (and hence the complementary polynomial $(\Im P, Q)$) can be constructed using the Weiss algorithm, the problem of finding phase factors can be solved by computing the inverse NLFT of the sequence $\bga$ such that $\overbrace{\bga}=(a,b)$. By the correspondence between QSP and NLFT in \cref{lem:HUdH-lemma-nonsym}, the phase factors can be obtained by setting $\psi_k=\arctan\gamma_k$ for $k=0,\ldots,d$.

\subsection{Layer stripping algorithm.}\label{sec:layer_stripping}

The \textit{layer stripping} algorithm to compute the inverse NLFT in the $\mathrm{SU}(2)$ case was introduced by Tsai in \cite{tsai2005nlft}, which follows the same idea as in the $\mathrm{SU}(1,1)$ case \cite{tao2012nonlinear}. The latter can be traced back to Schur's 1917 study using an algorithm now called the \textit{Schur algorithm}~\cite{schur1918potenzreihen}.  The layer stripping algorithm was developed independently in \cite{GilyenSuLowEtAl2019,Haah2019} for the purpose of finding phase factors in QSP, which is also called the \emph{peeling algorithm}.  The basic strategy is to strip off the unitary matrices one at a time from the left (or the right), thereby reducing the problem size by one each time, and then apply the method recursively to the smaller sequence until the entire sequence is read off, which gives the name ``layer stripping''.

We consider compactly supported sequences  $\bga \in \ell^\infty[(0,r)]$ for some $r \geq 1$. We will also use row vector notation to list the components of $\bga$, starting from index zero and up to some index $r' \ge r$. For example, if we write $\bga = [\gamma_m,\dots,\gamma_n]$, then it will mean that $\gamma_m$ is the component at index zero, and $\gamma_n$ is the component at index $r'$. The NLFT of $\bga$ will be denoted $\overbrace{[\gamma_m,\dots,\gamma_n]}$. For instance, with this notation, we have
\begin{equation}
    \overbrace{[\gamma_m,\ldots, \gamma_{n}]} = \overbrace{[\gamma_m,\ldots, \gamma_{n},0,0,\ldots]} = \prod_{j=0}^{n-m} \left[\frac{1}{\sqrt{1 + |\gamma_{j+m}|^2}} 
    \begin{pmatrix}
        1 & \gamma_{j+m} z^j \\
        - \ol{\gamma_{j+m}} z^{-j} & 1
    \end{pmatrix}\right].
\end{equation}

The problem of determining the first (left-most) component becomes finding $\gamma_0 \in \CC$ such that
\begin{equation}\label{eq: determine gamma_0}
    \frac{1}{\sqrt{1 + |\gamma_0|^2}} 
    \begin{pmatrix}
        1 & -\gamma_0 \\
        \ol{\gamma_0}  & 1
    \end{pmatrix}  \begin{pmatrix}
            a(z) & b(z)\\ -b^*(z) & a^*(z)
        \end{pmatrix} = \overbrace{[0,\gamma_1,\ldots, \gamma_{d}]}.
\end{equation}
If we let 
\begin{equation}\label{eq: a_1 b_1}
    \begin{pmatrix}
            a_1(z) &  b_1(z) \\ - b_1^*(z) & a_1^*(z)
        \end{pmatrix} = \overbrace{[\gamma_1,\ldots, \gamma_{d}]},
\end{equation} 
then \cref{eq: determine gamma_0} can be written as
\begin{equation}
\label{eq: determine a_1 and b_1}
    \frac{1}{\sqrt{1 + |\gamma_0|^2}}  \begin{pmatrix}
            a(z) + \gamma_0 b^\ast(z) & b(z) - \gamma_0 a^*(z)\\ \ol{\gamma_0} a(z) - b^*(z) & \ol{\gamma_0} b(z) + a^*(z)
        \end{pmatrix} = \begin{pmatrix}
            a_1(z) & z b_1(z) \\ -z^{-1} b_1^*(z) & a_1^*(z)
        \end{pmatrix}.
\end{equation}
Comparing the upper right element on both sides of the equation, the only way to make $\frac{b(z) - \gamma_0 a^*(z)}{z}$ be a polynomial is to let $\gamma_0 = \frac{b(0)}{a^*(0)}$, which is well defined since $a^\ast(0) > 0$. After determining $\gamma_0$, we can calculate $a_1(z)$ and $b_1(z)$ from \cref{eq: determine a_1 and b_1}. The remaining problem is to retrieve the rest of the sequence $\gamma_1, \gamma_2, \ldots, \gamma_{d-1}$ satisfying \cref{eq: a_1 b_1}. We may iteratively apply the same procedure to recover the remaining coefficients $\gamma_k$ one by one. 
The recursive formula takes the form
\begin{equation}
\label{eq: rec a_k^* b_k}
    \gamma_k = \frac{b_k(0)}{a_k^*(0)}, \quad a_{k+1}^*(z) = \frac{a_k^*(z) + \ol{\gamma_k} b_k(z)}{\sqrt{1 + |\gamma_k|^2}},\quad b_{k+1}(z) = \frac{b_k(z) - \gamma_k a_k^*(z)}{z \sqrt{1 + |\gamma_k|^2}}.
\end{equation}

The layer stripping algorithm is a sequential process. After determining $\gamma_k$, we need to compute $a_{k+1}^*$ and $b_{k+1}$ using \cref{eq: rec a_k^* b_k}, which requires $\Or(d-k)$ operations. Therefore, the overall complexity of the layer stripping algorithm is $\Or(d^2)$ operations.

\subsection{Riemann--Hilbert factorization algorithm.}\label{sec:RH_algorithm}

As discussed in \cref{sec:L2_conv_iqsp}, when $\bga \in \ell^2 (\ZZ)$ and is not compactly supported, the strategy is to split $\bga$ into two half-line supported sequences by solving the Riemann--Hilbert factorization problem in \cref{eq:factorization_explain}. After this, we may apply the layer stripping algorithm in \cref{sec:layer_stripping} to compute the phase factors of $\bga_-$ and $\bga_+$ separately, and then combine them to obtain the phase factors of $\bga$.

In fact, for any $k\in\ZZ$, there is a unique Riemann--Hilbert factorization that splits $\bga$ as $\bga_{k,-}+\bga_{k,+}$, with $\bga_{k,-}$ supported in $(-\infty, -k-1] $, $\bga_{k,+}$ supported in $[k,\infty)$, $a_{k,+}^*$ is outer, and $a_{k,+}^*(0)>0$~\cite[Theorem 5]{alexis2024infinite}. Here $\overbrace{\bga_{k,+}}=(a_{k,+},b_{k,+})$. We can just perform one step of the layer stripping algorithm on $(a_{k,+},b_{k,+})$  and obtain: 
\begin{equation}
\gamma_k = \frac{(b_{k,+} z^{-k}) (0)}{a_{k,+}^*(0)}.
\end{equation}
So for $(a,b)\in\mathcal{S}$, we can go through all $k$ in the support of $\bga$ and compute $\gamma_k$ one by one! This Riemann--Hilbert factorization is the only algorithm that can evaluate a single phase factor $\psi_k$ without computing all the other phase factors.

There are three steps to solve the Riemann--Hilbert factorization problem. Given the pair $(a,b)$ from the Weiss algorithm, we first compute the Laurent series
\begin{equation}
\frac{b(z)}{a(z)}=\sum_{j=-\infty}^d \hat{c}_j z^k.
\end{equation}
We will only need the coefficients $\hat{c}_0,\ldots,\hat{c}_d$, which can also be obtained from the Weiss algorithm using FFT~\cite[Algorithm 2]{alexis2024infinite}. Furthermore, all coefficients $\hat{c}_k$ are purely imaginary.

Second, we construct a Hankel matrix of size $(d-k+1)\times (d-k+1)$:
\begin{equation}
\Xi_k = 
\begin{pmatrix}
\hat{c}_k & \hat{c}_{k+1} & \cdots & \hat{c}_{d-1} & \hat{c}_d \\
\hat{c}_{k+1} & \hat{c}_{k+2} & \cdots & \hat{c}_d & 0 \\
\vdots & \vdots & \ddots & \vdots & \vdots \\
\hat{c}_{d-1} & \hat{c}_d & \cdots & 0 & 0 \\
\hat{c}_d & 0 & \cdots & 0 & 0
\end{pmatrix}, 
\end{equation}
and solve the linear system 
\begin{equation}\label{eqn:rh_linear_system}
\begin{pmatrix}
    I &  -\Xi_k\\
    -\Xi_k & I 
\end{pmatrix}\begin{pmatrix}
    \va_k\\
    \vb_k
\end{pmatrix} = \begin{pmatrix}
    \ve_0\\
    \bvec{0}
\end{pmatrix}
\end{equation}
for coefficients $\va_k$ and $\vb_k$. Here $\ve_0$ is the first column of the identity matrix.

Third, let $a_{k,0}$ and $b_{k,0}$ be the first entries of $\va_k$ and $\vb_k$. The result of the layer stripping on $(a_{k,+},b_{k,+})$ can be simply written as
\begin{equation}
\gamma_k = \frac{(b_{k,+} z^{-k}) (0)}{a_{k,+}^*(0)}=\frac{b_{k,0}}{a_{k,0}}.
\end{equation}

The complexity for solving the linear system in \cref{eqn:rh_linear_system} using standard methods is $\Or((d-k)^3)$ operations. Taking all $k$ in the support of $\bga$, the overall complexity is $\Or(d^4)$ operations. However, this is a pessimistic estimate since the linear systems for different $k$ are closely related and can be solved more efficiently. The half-Cholesky method in \cite{ni2024fast} reduces the computational cost to $\Or(d^2)$ operations. 

\subsection{Inverse nonlinear fast Fourier transform.}\label{sec:INLFFT}

Given $\bga$ supported on an interval of size $d$, the NLFT $\overbrace{\bga}=(a,b)$ can be computed in $\Or(d\log^2 d)$ operations using a \emph{nonlinear fast Fourier transform} (nonlinear FFT) algorithm developed in \cite{ni2024fast}. The basic idea is to use a divide-and-conquer strategy for fast polynomial multiplication, which is similar to the standard FFT algorithm. In this section, we discuss the \emph{inverse nonlinear fast Fourier transform} (inverse nonlinear FFT) algorithm~\cite{ni2025inverse}, which also relies on a divide-and-conquer structure and was inspired by the superfast Toeplitz system solver introduced in \cite{ammar1989numerical}.
Unlike the standard FFT, where the forward and inverse algorithms are nearly identical, the inverse nonlinear FFT differs substantially from its forward counterpart.

Let $m=\ceil{\frac{d+1}{2}}$. Recall the Riemann--Hilbert factorization at $m$ can be expressed as
\begin{equation}\label{eq: split seq}
    \overbrace{[\gamma_0,\ldots, \gamma_{m-1}]} \; \overbrace{[0,\ldots,0,\gamma_m,\ldots, \gamma_{d}]} = \overbrace{[\gamma_0,\ldots, \gamma_{d}]}.
\end{equation}
Let us define polynomials $\eta_{m}(z)$ and $\xi_{m}(z)$ by
\begin{equation}
    \overbrace{[\gamma_0,\ldots, \gamma_{m-1}]} = \begin{pmatrix}
        \eta_{m}^*(z)& \xi_{m}(z)\\
        -\xi_{m}^*(z)& \eta_{m}(z)
    \end{pmatrix}.
\end{equation}
In the matrix form, \cref{eq: split seq} becomes
\begin{equation}
    \begin{pmatrix}
        \eta_{m}^*(z)& \xi_{m}(z)\\
        -\xi_{m}^*(z)& \eta_{m}(z)
    \end{pmatrix}\begin{pmatrix}
            a_m(z) & z^m b_m(z)\\ -z^{-m} b_m^*(z) & a_m^*(z)
        \end{pmatrix} = \begin{pmatrix}
            a_0(z) & b_0(z)\\ -b_0^*(z) & a_0^*(z)
        \end{pmatrix}.
\end{equation}
We can invert the first matrix to obtain
\begin{equation}
\label{eq:key-layer-stripping-identity}
    \begin{pmatrix}
        z^m b_m(z)\\ a_m^*(z)
    \end{pmatrix} = \begin{pmatrix}
        \eta_{m}(z)& -\xi_{m}(z)\\
        \xi_{m}^*(z)& \eta_{m}^*(z)
    \end{pmatrix}\begin{pmatrix}
             b_0(z)\\ a_0^*(z)
        \end{pmatrix}.
\end{equation}
If $\eta_m,\xi_m$ are known, then $a_m,b_m$ can be computed using only a few fast polynomial multiplications with $\Or(d\log d)$ operations. 
Thus the remaining task is to compute $\eta_m(z)$ and $\xi_m(z)$ from $[\gamma_0,\ldots, \gamma_{m-1}]$ efficiently, which can be obtained in a recursive fashion.  Specifically, let $l = \ceil{\frac{m}{2}}$, and assume that the recursive steps have already yielded
\begin{equation}
\overbrace{[\gamma_0,\ldots, \gamma_{l-1}]} \quad \text{and}\quad \overbrace{[\gamma_l,\ldots, \gamma_{m-1}]},
\end{equation}
then we may compute $\overbrace{[\gamma_0,\ldots, \gamma_{m-1}]}$ using \cref{eq: split seq} and fast polynomial multiplications. We refer readers to~\cite[Algorithm 1]{ni2025inverse} for details of the inverse nonlinear FFT algorithm.  The total computational complexity is only
\begin{equation}
    d\log d + 2\left(\frac{d}{2}\log\frac{d}{2}\right) + 4\left(\frac{d}{4}\log\frac{d}{4}\right) + \cdots = \Or(d\log^2 d).
\end{equation}
Since any algorithm needs to read all $d$ components of $\bga$, the optimal complexity is $\Or(d)$. Thus the inverse nonlinear FFT algorithm achieves the \emph{near optimal} complexity of $\Or(d\log^2 d)$ operations.

\section{Numerical stability analysis.}\label{sec:numerical_stability}

\subsection{Floating point arithmetic and error propagation.}\label{sec:floating-point-background}

We briefly review some fundamental concepts of numerical error propagation \cite{Higham2002}. The standard model of floating-point arithmetic states that for any basic arithmetic operation $\circ$, the computed result $\mathrm{fl}(a \circ b)$ satisfies
\begin{equation}
\mathrm{fl}(a \circ b) = (a \circ b)(1 + \delta), \quad |\delta| \leq \epsilon_{\mathrm{ma}},
\end{equation}
where $\epsilon_{\mathrm{ma}}$ denotes the machine precision.

Consider an algorithm whose exact output is $\vx$, and suppose the target precision is $\epsilon$. We say the algorithm has a bit requirement $r$ if, when using floating point arithmetic with $\epsilon_{\mathrm{ma}} = 2^{-r}$, the computed output $\hat{\vx}$ satisfies $\frac{\|\vx - \hat{\vx}\|}{\norm{\vx}} \leq \epsilon$. In practice, we prefer algorithms that operate reliably under fixed precision regardless of the problem size, such as $r = 52$ for IEEE double precision. However, in the worst case, numerical error may accumulate, making this goal theoretically unattainable. We say an algorithm is \emph{numerically stable} if the bit requirement is $r = \Or(\operatorname{polylog}(d, 1/\epsilon))$, where $d$ denotes the problem size. In practice, such algorithms often perform robustly using standard double precision arithmetic operations. On the other hand, an algorithm is numerically unstable if the bit requirement is $r = \Omega(\poly(d))$, in which case the error can accumulate rapidly for moderate values of $d$ in practice.

Numerical stability is typically assessed via forward and backward error analysis. The forward error measures the deviation of the computed solution $\hat{\vx}$ from the exact solution $\vx$, while the backward error reflects the smallest perturbation to the input that would make $\hat{\vx}$ an exact solution. For example, when solving a linear system $A \vx = \vb$, assuming $A, \vb$ are non-zero, we have ($\norm{A}$ denotes the operator norm induced by the vector 2-norm $\norm{\cdot}$):
\begin{align}
\text{Forward (relative) error } &:= \frac{\|\hat{\vx} - \vx\|}{\|\vx\|}, \\
\text{Backward (relative) error } &:= \min_{\Delta A} \left\{ \frac{\|\Delta A\|}{\|A\|} : (A + \Delta A)\hat{\vx} = \vb \right\},
\end{align}
For linear systems, forward and backward errors are linked by the condition number $\kappa(A):= \|A\| \|A^{-1}\|$, with the forward error bounded by the product of the condition number and the backward error \cite[Theorem 7.2]{Higham2002}.

\subsection{Numerical stability of Weiss algorithm.}

The Weiss algorithm is based on the Hilbert transform, which can be computed using FFT, which is a numerically stable procedure. The main source of the difficulty arises when $\sup\limits_{z \in \TT} |b(z)|=1-\eta$ and $\eta$ is very small, and as a result the magnitude of the function $\log(1-\abs{b}^2)$ on $\TT$ becomes large. The Weiss algorithm requires $\mathcal{O}(\log(\frac{d}{\epsilon \eta}))$ bits of precision~\cite[Section 5.5]{alexis2024infinite}. Thus it is a numerically stable algorithm.

\subsection{Gaussian elimination, displacement structure, and numerical stability of layer stripping algorithm.}

Haah's analysis~\cite{Haah2019} showed that the layer stripping algorithm described in \cref{sec:layer_stripping} also requires $\Or(d\log(d/\epsilon))$ bits of precision, and is thus \emph{numerically unstable}. This is because even small errors can accumulate exponentially during the recursive process of the layer stripping algorithm, as can be observed from numerical experiments. 

Is there a set of sufficient conditions that guarantee numerical stability of inverse NLFT?  Ref.~\cite{ni2025inverse} showed that when $a^*$ is an outer function, the layer stripping algorithm is in fact numerically stable. The proof is based on the connection between the layer stripping algorithm and Schur algorithm, which is in turn related to the Gaussian elimination process for matrices with \emph{displacement low-rank structure}~\cite{KailathSayed1995}. For a matrix $A$, its conjugate transpose is denoted by $A^{\dagger}$.

A matrix $K$ is said to have displacement rank $r$ if the matrix $K - Z_nKZ_n^{\dag}$ is of rank $r$, where the lower shift matrix is
\begin{equation}
    Z_n:= \begin{pmatrix}
    0&&&\\
    1&0&&\\
    &\ddots &\ddots&\\
    &&1&0
\end{pmatrix}_{n\times n}.
\end{equation}
Let $n=d+1$. In the layer stripping algorithm, let $b_k(z) := \sum_{j=0}^{d-k} b_{j,k} z^j$ and $a_k^*(z) := \sum_{j=0}^{d-k} a_{j,k} z^j$.  Note that we are using the coefficients of $a_k^*(z)$ instead of $a_k(z)$ to avoid negative powers of $z$. We also point out that $(a_0,b_0) = (a,b)$ is the input pair. Define the column vectors $\ba_k := (a_{j,k})_{0\le j\le d-k}$, $\bb_k := (b_{j,k})_{0\le j\le d-k}$.  Also define a matrix $K := T(\ba_0)T(\ba_0)^{\dag} + T(\bb_0)T(\bb_0)^{\dag}$, where $T(\ba_0)$ is the lower triangular Toeplitz matrix
\begin{equation}
\label{eq:Ta0-def}
    T(\ba_0) := \begin{pmatrix}
        a_{0,0}&&&&\\
        a_{1,0}&a_{0,0}&&&\\
        a_{2,0}&a_{1,0}&a_{0,0}&&\\
        \vdots&\ddots&\ddots&\ddots&\\
        a_{d,0}&\cdots&a_{2,0}&a_{1,0}&a_{0,0}
    \end{pmatrix},
\end{equation}
with first column $\ba_0 = (a_{0,0}, a_{1,0},\ldots,a_{d,0})^T$.
It follows from \cite[Lemma~2]{KailathSayed1995} that $K$ satisfies 
\begin{equation}\label{eq: displacement structure K}
    K - Z_{d+1} K Z_{d+1}^{\dag} = \ba_0\ba_0^{\dag} + \bb_0\bb_0^{\dag},
\end{equation}
i.e., $K$ has displacement rank 2. Furthermore, the layer stripping algorithm is equivalent to performing Gaussian elimination (in fact the Cholesky factorization) on $K$. At the end of the layer stripping algorithm, we have $K=LDL^{\dag}$ where $L$ is a unit lower triangular matrix and $D$ is a diagonal matrix with positive entries. The significance of relating layer stripping with the displacement structure is that the Cholesky factorization is a backward stable algorithm for positive definite matrices~\cite[Chapter 10]{Higham2002}. The algorithm is forward stable if the diagonal entries of $D$ are not too small. In this case, the forward stability can be guaranteed when $\sup\limits_{z \in \TT} |b(z)|=1-\eta$ for any $\eta>0$ and choosing $a^*$ to be an outer function. We can prove a stronger result that the condition number of $K$ is bounded by $\Or(1/\eta)$~\cite[Lemma 5.4]{ni2025inverse}, and refer readers to~\cite[Section 5.4]{ni2025inverse}.

\subsection{Numerical stability of Riemann--Hilbert factorization.}

The Riemann--Hilbert factorization algorithm was the first provably numerically stable algorithm for finding phase factors. The proof of the numerical stability is in fact simpler than that of the layer stripping algorithm, since all phase factors can be computed independently and there is no recursive process involved. 

Recall that in the linear system~\cref{eqn:rh_linear_system}, when $a^*$ is outer, all entries of $\Xi_k$ are purely imaginary. Thus the smallest singular value of the coefficient matrix is at least $1$, and the condition number of the coefficient matrix can thus be bounded. The numerical stability of the Riemann--Hilbert factorization algorithm then follows from the backward stability of the Gaussian elimination process (in fact, Cholesky factorization) of an equivalent positive definite system~\cite[Section 5.5]{alexis2024infinite}.

\subsection{Numerical stability of inverse nonlinear fast Fourier transform.}

The proof of the numerical stability of the inverse nonlinear FFT algorithm follows a structure similar to that of the layer stripping algorithm. In particular, it can be viewed as a fast algorithm to factorize the matrix $K$, and when $a^*$ is outer, the condition number of $K$ is bounded~\cite[Lemma 5.4]{ni2025inverse}. However, the backward stability analysis of the inverse nonlinear FFT algorithm is much more involved due to the recursive nature of the algorithm. We refer readers to~\cite[Section 5.5]{ni2025inverse}.

\section{Iterative algorithms for finding phase factors.}\label{sec:iterative_algorithms}

Let us view the QSP phase factor finding problem from a different angle. Given a target polynomial $f\in\RR[x]$ of degree $d$ satisfying (1) the parity of $f$ is $d \bmod 2$, and (2) $\norm{f}_{\infty}\le 1$, we want to find phase factors $\Psi\in\RR^{d+1}$ such that $f(x)$ is equal to the real (or imaginary) part of the top-left entry of $U_d(x,\Psi)$ for all $x\in[-1,1]$. 
The mapping from the target polynomial of degree $d$ (described by its Chebyshev coefficients denoted by $\vc\in\RR^{d+1}$) to phase factors $\Psi\in\RR^{d+1}$ can be abstractly written as
\begin{equation}\label{eq:obj}
    F(\Psi)=\vc.
\end{equation}
The mapping $F$ is highly nonlinear and is not one-to-one. For a given $\vc$, our goal is to find \emph{one} solution to the nonlinear system \eqref{eq:obj}. This can be also viewed as an optimization problem
\begin{equation}\label{eqn:optimization}
    \Psi^* = \argmin_{\Psi} \norm{F(\Psi) - \vc}_2^2.
\end{equation}
Since a target polynomial can be exactly represented by phase factors, the minimum value of the optimization problem is zero. However, due to the complex energy landscape~\cite{WangDongLin2022}, direct optimization from random initial guesses can easily get stuck at local minima and can only be used for low degree polynomials.  Ref.~\cite{DongMengWhaleyEtAl2021} observed that starting from the same, problem-independent initial phase factors $\Psi^0 = (0,0,\ldots,0)$, 
standard optimization (such as gradient or quasi-Newton type) methods can be used to robustly evaluate the phase factors.

The main advantages of iterative algorithms are their simplicity and numerical stability. It does not require the construction of complementary polynomials. Each iteration only requires the evaluation of $F(\Psi)$ and its Jacobian $J(\Psi)$, which only involves matrix multiplications and is numerically stable. 
This leads to the first practical algorithm to find symmetric phase factors for polynomials of degree up to a few thousands. From a theoretical perspective, so far it is only known that when the target function is scaled as $\norm{f}_\infty = \Or(1/d)$, the optimization-based algorithm converges locally, and the computational cost is $\Or(d^2 \log(1/\epsilon))$~\cite{WangDongLin2022}.

The simplest iterative method to solve \cref{eq:obj}, and perhaps the simplest algorithm among all algorithms for finding phase factors, is the fixed point iteration (FPI) algorithm introduced in~\cite{DongLinNiEtAl2024_iqsp}, which consists of only a single line:
\begin{equation}\label{eq:FPI_update}
\Phi^{0} = \mathbf{0}\in\RR^{\wt{d}},\quad\Phi^{t+1}=\Phi^{t}-\frac{1}{2}\left( F\left(\Phi^{t}\right)-\vc\right).
\end{equation}
Here $\wt{d}=\lceil \frac{d+1}{2} \rceil$ is the number of symmetric phase factors.  $\Phi^t$ is the column vector for the reduced phase factors at the $t$-th iteration, and $c\in\RR^{\wt{d}}$ is the target Chebyshev coefficients. The FPI algorithm converges linearly to the maximal solution when $\norm{c}_1\le 0.861$ based on a contraction mapping argument~\cite{DongLinNiEtAl2024_iqsp}. The computational complexity of the FPI algorithm is $\Or(d^2)$ operations per iteration, and can be reduced to 
$\Or(d\log^2 d)$ operations using fast polynomial multiplication~\cite{ni2024fast}. 
 Numerical experiments show that the FPI algorithm can be used to find phase factors for polynomials of degree up to a few thousands when $\norm{c}_1$ is close to $1$.

Furthermore, \cite{DongLinNiEtAl2024_newton} proposed a Newton-type algorithm to solve \cref{eq:obj}, which is empirically observed to converge rapidly and robustly in all parameter regimes starting from $\Phi^{0} = \mathbf{0}$. However, the cost of each iteration increases to $\Or(d^3)$, which becomes significant for large problems.
It remains an open question to establish the theoretical guarantees to justify the superior performance of the Newton-type algorithm.

\section{Quantum singular value transformation and its applications.}\label{sec:qsvt}


So far we have introduced in detail the mathematical structure of QSP and various algorithms for determining phase factors. Quantum singular value transformation (QSVT)~\cite{GilyenSuLowEtAl2019} can be regarded as a natural generalization of QSP: while QSP acts on scalars, QSVT encodes polynomial transformations of singular values of matrices, which can then represent a versatile set of matrix-function transformations.
It has since been recognized as a seminal development in quantum algorithms, and provides a unifying framework for many existing and new quantum algorithms \cite{MartynRossiTanEtAl2021}.  

Just like we embed a scalar $x$ into a $2\times 2$ unitary matrix $W(x)$ in \cref{eqn:qsp-unitary}, when we are given a matrix $A\in\CC^{N\times N}$ with singular values in the interval $[0,1]$, we can embed it into a unitary matrix $U_A$, called a \emph{block encoding} of $A$. Here we consider the simplest case, where $A$ is embedded into a $2N\times 2N$ unitary matrix
\begin{equation*}
U_A=\begin{pmatrix}
{A} & {*} \\
{*} & {*}
\end{pmatrix}
\end{equation*}
where each matrix block $*$ is an $N\times N$ matrix. Their values are irrelevant for the task of QSVT as long as $U_A$ is unitary. In practice, $U_A$ should be efficiently implemented on quantum computers. For instance, when $A$ is a sparse matrix, $U_A$ can be implemented efficiently using oracles that encode the locations and values of nonzero entries~\cite{BerryChildsCleveEtAl2014,GilyenSuLowEtAl2019}. We will not get into the details here.

The definition of QSVT depends on the parity of the target function $f$, assumed to be an even or odd polynomial of degree $d$ here for simplicity. Let the singular value decomposition of $A$ be $A=W\Sigma V^\dagger$, where $\Sigma=\operatorname{diag}(\sigma_0,\ldots,\sigma_{N-1})$ with singular values $\sigma_i\in[0,1]$, and $V^{\dagger}$ is the conjugate transpose of $V$. Then the singular value transformation of $A$ is defined as
\begin{equation}
f^{\mathrm{SV}}(A)=\begin{cases}
W f(\Sigma) V^{\dag},& f \mbox{ is odd},\\
V f(\Sigma) V^{\dag},& f \mbox{ is even},\\
\end{cases} 
\qquad f\left(\Sigma\right)=\operatorname{diag}\left(f\left(\sigma_{0}\right), f\left(\sigma_{1}\right), \ldots, f\left(\sigma_{N-1}\right)\right).
\end{equation}
Note that when $A$ is a Hermitian matrix, the singular value transformation is equivalent to the standard functional calculus of matrices $f(A)$. 

Then QSVT provides an elegant construction of a unitary matrix $U_f\in\CC^{2N\times 2N}$ such that
\begin{equation}
U_f=\begin{pmatrix}
f^{\mathrm{SV}}(A) & {*} \\
{*} & {*}
\end{pmatrix},
\end{equation}
using $d$ queries to the unitaries $U_A$ or $U_A^\dagger$, interleaved by $d+1$ single qubit rotations parameterized by the phase factors $\Psi\in\RR^{d+1}$ corresponding to $f$. In other words, QSVT generalizes QSP from scalars to matrices, by constructing a block encoding of $f^{\mathrm{SV}}(A)$. This ``lifting'' procedure from scalars to matrices is called \emph{qubitization}. We refer interested readers to Refs.~\cite{low2019hamiltonian,GilyenSuLowEtAl2019,MartynRossiTanEtAl2021}. It is also worth noting that qubitization can be compactly viewed~\cite{Dong2023thesis,tang2024cs} from the perspective of the \emph{cosine-sine (CS) decomposition}~\cite{paige1994history} in linear algebra. 

QSVT has found many applications in quantum algorithms,such as Hamiltonian simulation~\cite{LowChuang2017,GilyenSuLowEtAl2019},  linear system of equations~\cite{GilyenSuLowEtAl2019,LinTong2020}, eigenvalue problems~\cite{LinTong2020a,DongLinTong2022}, solving differential equations~\cite{FangLinTong2023}, Petz recovery channel~\cite{GilyenLloydMarvianEtAl2022}, to name a few. Here are a few examples:
\begin{enumerate}
\item Hamiltonian simulation: Given a Hermitian matrix $H\in\CC^{N\times N}$,  with $\norm{H}\le 1$ and $t>0$, construct a block encoding for $e^{-iHt}$. This can be achieved by constructing a block encoding of $\cos(Ht)$ and $\sin(Ht)$ separately using QSVT. Specifically, 
we first use the Fourier--Chebyshev series of the trigonometric functions on $[-1,1]$ (called the Jacobi-Anger expansion):
\begin{equation}
\cos (t x)=J_{0}(t)+2 \sum_{k=1}^{\infty}(-1)^{k} J_{2 k}(t) T_{2 k}(x), \quad
\sin (t x)=2 \sum_{k=0}^{\infty}(-1)^{k} J_{2 k+1}(t) T_{2 k+1}(x).
\label{eqn:jacobi_anger}
\end{equation}
Here $J_{\nu}(t)$ denotes Bessel functions of the first kind.
This series converges very rapidly, and the number of terms needed to approximate $\cos(tx)$ and $\sin(tx)$ with uniform error $\epsilon$ on $[-1,1]$ is $\Or(t+\log(1/\epsilon))$. This gives rise to a Hamiltonian simulation algorithm with asymptotically optimal scaling~\cite{LowChuang2017,GilyenSuLowEtAl2019}. It is also worth noting that the original QSP representation~\cite{LowChuang2017} queries a ``quantum walk'' oracle rather than the block encoding oracle. 

\item Solving linear systems of equations: 
Given an invertible matrix $A\in\CC^{N\times N}$ with singular values in $[\kappa^{-1},1]$, a key step in solving the linear system of equations $Ax=b$ on a quantum computer is to 
construct a block encoding of $A^{-1}/\kappa$.
From the SVD $A=W\Sigma V^{\dag}$, the matrix inverse can be expressed as (note that $V^{\dag}=V^{-1}$ since $V$ is unitary)
\begin{equation}
A^{-1}/\kappa=V (\kappa\Sigma)^{-1} W^\dagger=f^{\mathrm{SV}}(A^\dagger),
\label{eqn:ainv_qsvt}
\end{equation}
where $f(x)=(\kappa x)^{-1}$ is an odd function,  and $f(x)$ can be approximated by an odd polynomial of degree $\Or(\kappa\log(1/\epsilon))$ with uniform error $\epsilon$ on $[-1,-\kappa^{-1}]\cup[\kappa^{-1},1]$~\cite{ChildsKothariSomma2017,GilyenSuLowEtAl2019}. The desired block encoding can be constructed by applying QSVT to $A^{\dag}$.

\item Eigenvalue problems:
Given a Hermitian matrix $H\in\CC^{N\times N}$ with eigenvalues in $[0,1]$, and a real number $x_0\in[0,1]$, we want to construct a block encoding of the spectral projector $\Pi_{H\le x_0}$ that projects onto the eigenspace of $H$ with eigenvalues less than or equal to $x_0$, with the guarantee that there are no eigenvalues in the interval $(x_0-\delta,x_0+\delta)$ for some $\delta\in(0,1)$. Since all eigenvalues of $H$ are non-negative, this can be achieved by constructing an even approximation to the step function
\begin{equation}
f(x)=\begin{cases}
1,& |x| < x_0,\\
0,& |x| > x_0,
\end{cases}
\end{equation}
with uniform error $\epsilon$ on $[0,x_0-\delta]\cup[x_0+\delta,1]$, and the polynomial degree is $\Or(\frac{\log(1/\epsilon)}{\delta})$ ~\cite{LowChuang2017a,GilyenSuLowEtAl2019}. This is a key step for the near optimal algorithm for solving eigenvalue problems~\cite{LinTong2020a} and for solving linear systems of equations~\cite{LinTong2020} using QSVT.



\end{enumerate}

\section{Generalizations of quantum signal processing and outlook.}\label{sec:generalizations_outlook}

In the standard QSP representation, each parameterized unitary $e^{i \psi_k Z}$ only has one (real) degree of freedom. We may also parameterize each unitary by two angles as
\begin{equation}
\label{eq:R-def}
    R(\psi,\phi) := \begin{pmatrix}
        \cos\psi& e^{i\phi}\sin\psi\\ 
        -e^{-i\phi}\sin\psi & \cos\psi
    \end{pmatrix}, \;\;\; \psi,\phi \in [-\pi,\pi],
\end{equation}
This is a slight variation of the \emph{generalized quantum signal processing} (GQSP) task proposed in~\cite{MotlaghWiebe2024}. Specifically,  given a target polynomial $b(z)\in\CC[z]$ of degree $d$ satisfying $\sup\limits_{z \in \TT} |b(z)| \le 1$, GQSP seeks to find sequences $\{\phi_k\}_{k=0}^d$ and $\{\psi_k\}_{k=0}^d$ such that
\begin{equation}\label{eq:gqsp-def}
\begin{pmatrix}
    \cdot & b(z)\\ \cdot & \cdot
\end{pmatrix} = R(\psi_0,\phi_0) \prod_{k=1}^d \left(\begin{pmatrix}
    z&\\&1
\end{pmatrix}R(\psi_k,\phi_k)\right).
\end{equation}
Recall that in QSP, $f$ can be a real (or imaginary) polynomial. By writing $f(x)=f((z+z^{-1})/2)$ with $z\in\TT$, we see that $f(z)$ is a Laurent polynomial of degree $d$ that must satisfy a parity constraint. In GQSP, $b(z)$ can be a general complex polynomial of degree $d$ without parity constraints. However, $b(z)$ can only be a polynomial (i.e., analytic function), not a Laurent polynomial. The existence of the phase factors $\{\phi_k\}$ and $\{\psi_k\}$ is guaranteed by \cite[Corollary 5]{MotlaghWiebe2024}. The quantum eigenvalue transformation
of unitary matrices with real polynomials (QETU)~\cite{DongLinTong2022} can be mapped to a GQSP problem after choosing special phase angles for $\{\phi_k\}$. Following a construction similar to QSVT,
GQSP and QETU can be used to construct a block encoding of $f(H)$ for a Hermitian matrix $H$, by querying the Hamiltonian evolution $e^{iHt}$ instead of a block encoding of $H$; see also~\cite{sunderhauf2023generalized} for a generalization to block encoding query models. 

\cite[Theorem 3.3]{ni2025inverse} shows that the GQSP problem and NLFT problem are equivalent. In particular, given a target polynomial $b(z)$ that can be expressed as $\overbrace{\bga}=(a,b)$, then the
corresponding GQSP phase factor sequences are determined by $\psi_k = \arctan(|\gamma_k|)$ and $\phi_k = \Arg(\gamma_k)$, for $k=0,\dots,d$. Such a connection between GQSP and NLFT also appears in \cite{laneve2025generalized}.

There are several other generalizations of quantum signal processing, including $\mathrm{SU}(1,1)$ (also called the continuous variable setting)~\cite{RossiBastidasMunroEtAl2023,liu2024hybrid,singh2025towards}, $\mathrm{SU}(N) $~\cite{laneve2023quantum,lu2024quantum}, parallel QSP~\cite{martyn2025parallel}, and multi-variable QSPs~\cite{RossiChuang2022,gomes2024multivariable,laneve2025generalized}. Compared to the univariate case, the characterization of achievable polynomials and the corresponding algorithms for determining the parameters are much less developed in the multivariate setting. While QSP-type representations may universally approximate a multivariate continuous function $f:[0,1]^m \to \mathbb{C}$~\cite[Theorem 4]{perez2021one}, such representations are not constructive. Moreover, in the multivariate setting, there exist polynomial pairs $(P, Q)$ that do not admit a QSP type decomposition~\cite{nemeth2023variants,laneve2025multivariate}; see also~\cite{laneve2025adversary} for constraints on the class of achievable polynomials. Thus the complementary polynomials may play an even more important role in the multivariate case, and perspectives from nonlinear Fourier analysis may be useful in addressing these challenges. Another significant challenge is that so far there is no analog of QSVT that can be used to lift these generalizations of QSP from scalars to matrices. If such a lifting can be achieved, it may lead to a new class of quantum algorithms with new applications in scientific computation.

\subsection*{Acknowledgments.}\label{ssec:acknowledgment}
 This work is partially supported by the Challenge Institute for Quantum Computation (CIQC) funded by the National Science Foundation (NSF) through Grant No. OMA-2016245, by
the U.S. Department of Energy, Office of Science, Office of Advanced Scientific Computing Research's Applied Mathematics Competitive Portfolios program under Contract No. AC02-05CH11231, and by a Simons Investigator award through Grant Number 825053.  We thank Michel Alexis, Yulong Dong, Lorenzo Laneve, James Larsen, Yuan Liu, Guang Hao Low, John Martyn, Gevorg Mnatsakanyan, Hongkang Ni, Zane Rossi, Rahul Sarkar, Christoph Thiele, Jiasu Wang, Lexing Ying for collaborations on related projects and helpful comments on the manuscript.


\begin{thebibliography}{10}

\bibitem{alexis2024infinite}
M.~Alexis, L.~Lin, G.~Mnatsakanyan, C.~Thiele, and J.~Wang.
\newblock Infinite quantum signal processing for arbitrary {S}zeg{\H o}
  functions.
\newblock {\em Commun. Pure Appl. Math. in press}, 2025.

\bibitem{alexis2024quantum}
M.~Alexis, G.~Mnatsakanyan, and C.~Thiele.
\newblock Quantum signal processing and nonlinear {F}ourier analysis.
\newblock {\em Revista Matem{\'a}tica Complutense}, 37:1--40, 2024.

\bibitem{ammar1989numerical}
G.~S. Ammar and W.~B. Gragg.
\newblock Numerical experience with a superfast real {T}oeplitz solver.
\newblock {\em Linear Algebra Appl.}, 121:185--206, 1989.

\bibitem{beals1985inverse}
R.~Beals and R.~R. Coifman.
\newblock Inverse scattering and evolution equations.
\newblock {\em Commun. Pure Appl. Math.}, 38:29--42, 1985.

\bibitem{berntson2025complementary}
B.~K. Berntson and C.~S{\"u}nderhauf.
\newblock Complementary polynomials in quantum signal processing.
\newblock {\em Commun. Math. Phys.}, 406:161, 2025.

\bibitem{BerryChildsCleveEtAl2014}
D.~W. Berry, A.~M. Childs, R.~Cleve, R.~Kothari, and R.~D. Somma.
\newblock Exponential improvement in precision for simulating sparse
  {H}amiltonians.
\newblock In {\em Proceedings of the forty-sixth annual ACM symposium on Theory
  of computing}, pages 283--292, 2014.

\bibitem{case1975orthogonal}
K.~M. Case.
\newblock Orthogonal polynomials. {II}.
\newblock {\em J. Math. Phys.}, 16:1435--1440, 1975.

\bibitem{ChildsKothariSomma2017}
A.~M. Childs, R.~Kothari, and R.~D. Somma.
\newblock Quantum algorithm for systems of linear equations with exponentially
  improved dependence on precision.
\newblock {\em SIAM J. Comput.}, 46:1920--1950, 2017.

\bibitem{damanik2004half}
D.~Damanik and R.~Killip.
\newblock Half-line {S}chr{\"o}dinger operators with no bound states.
\newblock {\em Acta Math.}, 193:31--72, 2004.

\bibitem{deift2000orthogonal}
P.~A. Deift.
\newblock {\em Orthogonal polynomials and random matrices: a
  {R}iemann-{H}ilbert approach}, volume~3.
\newblock American Mathematical Society, 2000.

\bibitem{denisov2002probability}
S.~A. Denisov.
\newblock Probability measures with reflection coefficients $a_n \in \ell^4$
  and $a_{n+ 1}- a_n \in \ell^2$ are {E}rd{\"o}s measures.
\newblock {\em Journal of Approximation Theory}, 117:42--54, 2002.

\bibitem{Dong2023thesis}
Y.~Dong.
\newblock {\em Quantum signal processing algorithm and its applications}.
\newblock PhD thesis, University of California, Berkeley, 2023.

\bibitem{DongLinNiEtAl2024_iqsp}
Y.~Dong, L.~Lin, H.~Ni, and J.~Wang.
\newblock Infinite quantum signal processing.
\newblock {\em Quantum}, 8:1558, 2024.

\bibitem{DongLinNiEtAl2024_newton}
Y.~Dong, L.~Lin, H.~Ni, and J.~Wang.
\newblock Robust iterative method for symmetric quantum signal processing in
  all parameter regimes.
\newblock {\em SIAM J. Sci. Comput.}, 46:A2951--A2971, 2024.

\bibitem{DongLinTong2022}
Y.~Dong, L.~Lin, and Y.~Tong.
\newblock Ground-state preparation and energy estimation on early
  fault-tolerant quantum computers via quantum eigenvalue transformation of
  unitary matrices.
\newblock {\em PRX Quantum}, 3:040305, 2022.

\bibitem{DongMengWhaleyEtAl2021}
Y.~Dong, X.~Meng, K.~B. Whaley, and L.~Lin.
\newblock Efficient phase factor evaluation in quantum signal processing.
\newblock {\em Phys. Rev. A}, 103:042419, 2021.

\bibitem{dym2008gaussian}
H.~Dym and H.~P. McKean.
\newblock {\em Gaussian processes, function theory, and the inverse spectral
  problem}.
\newblock Dover Publications, 2008.

\bibitem{faddeev1987hamiltonian}
L.~Faddeev, A.~Reyman, and L.~Takhtajan.
\newblock {\em Hamiltonian Methods in the Theory of Solitons}.
\newblock Springer Berlin Heidelberg, 2007.

\bibitem{FangLinTong2023}
D.~Fang, L.~Lin, and Y.~Tong.
\newblock Time-marching based quantum solvers for time-dependent linear
  differential equations.
\newblock {\em Quantum}, 7:955, 2023.

\bibitem{fokas1994integrability}
A.~S. Fokas and I.~Gelfand.
\newblock Integrability of linear and nonlinear evolution equations and the
  associated nonlinear {F}ourier transforms.
\newblock {\em Lett. Math. Phys.}, 32:189--210, 1994.

\bibitem{GilyenLloydMarvianEtAl2022}
A.~Gily{\'e}n, S.~Lloyd, I.~Marvian, Y.~Quek, and M.~M. Wilde.
\newblock Quantum algorithm for petz recovery channels and pretty good
  measurements.
\newblock {\em Phys. Rev. Lett.}, 128(22):220502, 2022.

\bibitem{GilyenSuLowEtAl2019}
A.~Gily{\'e}n, Y.~Su, G.~H. Low, and N.~Wiebe.
\newblock Quantum singular value transformation and beyond: exponential
  improvements for quantum matrix arithmetics.
\newblock In {\em Proceedings of the 51st Annual ACM SIGACT Symposium on Theory
  of Computing}, pages 193--204, 2019.

\bibitem{gomes2024multivariable}
N.~Gomes, H.~Lim, and N.~Wiebe.
\newblock Multivariable {QSP} and bosonic quantum simulation using iterated
  quantum signal processing.
\newblock {\em arXiv preprint arXiv:2408.03254}, 2024.

\bibitem{Haah2019}
J.~Haah.
\newblock Product decomposition of periodic functions in quantum signal
  processing.
\newblock {\em Quantum}, 3:190, 2019.

\bibitem{Higham2002}
N.~J. Higham.
\newblock {\em Accuracy and stability of numerical algorithms}, volume~80.
\newblock SIAM, 2002.

\bibitem{hitrik2001properties}
M.~Hitrik.
\newblock Properties of the scattering transform on the real line.
\newblock {\em Journal of Mathematical Analysis and Applications},
  258:223--243, 2001.

\bibitem{KailathSayed1995}
T.~Kailath and A.~H. Sayed.
\newblock Displacement structure: Theory and applications.
\newblock {\em SIAM Rev.}, 37:297--386, 1995.

\bibitem{killip2003sum}
R.~Killip and B.~Simon.
\newblock Sum rules for {J}acobi matrices and their applications to spectral
  theory.
\newblock {\em Ann. Math.}, 158:253--321, 2003.

\bibitem{koosis1998logarithmic}
P.~Koosis.
\newblock {\em The Logarithmic Integral {I}}, volume~1.
\newblock Cambridge University Press, 1998.

\bibitem{laneve2023quantum}
L.~Laneve.
\newblock Quantum signal processing over {SU(N)}.
\newblock {\em arXiv preprint arXiv:2311.03949}, 2023.

\bibitem{laneve2025adversary}
L.~Laneve.
\newblock An adversary bound for quantum signal processing.
\newblock {\em arXiv preprint arXiv:2506.20484}, 2025.

\bibitem{laneve2025generalized}
L.~Laneve.
\newblock Generalized quantum signal processing and non-linear {F}ourier
  transform are equivalent.
\newblock {\em arXiv preprint arXiv:2503.03026}, 2025.

\bibitem{laneve2025multivariate}
L.~Laneve and S.~Wolf.
\newblock On multivariate polynomials achievable with quantum signal
  processing.
\newblock {\em Quantum}, 9:1641, 2025.

\bibitem{LinTong2020a}
L.~Lin and Y.~Tong.
\newblock Near-optimal ground state preparation.
\newblock {\em Quantum}, 4:372, 2020.

\bibitem{LinTong2020}
L.~Lin and Y.~Tong.
\newblock Optimal quantum eigenstate filtering with application to solving
  quantum linear systems.
\newblock {\em Quantum}, 4:361, 2020.

\bibitem{liu2024hybrid}
Y.~Liu, S.~Singh, K.~C. Smith, E.~Crane, J.~M. Martyn, A.~Eickbusch,
  A.~Schuckert, R.~D. Li, J.~Sinanan-Singh, M.~B. Soley, et~al.
\newblock Hybrid oscillator-qubit quantum processors: Instruction set
  architectures, abstract machine models, and applications.
\newblock {\em arXiv preprint arXiv:2407.10381}, 2024.

\bibitem{LowChuang2017a}
G.~H. Low and I.~L. Chuang.
\newblock Hamiltonian simulation by uniform spectral amplification.
\newblock {\em arXiv:1707.05391}, 2017.

\bibitem{LowChuang2017}
G.~H. Low and I.~L. Chuang.
\newblock Optimal {H}amiltonian simulation by quantum signal processing.
\newblock {\em Phys. Rev. Lett.}, 118:010501, 2017.

\bibitem{low2019hamiltonian}
G.~H. Low and I.~L. Chuang.
\newblock Hamiltonian simulation by qubitization.
\newblock {\em Quantum}, 3:163, 2019.

\bibitem{lu2024quantum}
X.~Lu, Y.~Liu, and H.~Lin.
\newblock Quantum signal processing and quantum singular value transformation
  on {U(N)}.
\newblock {\em arXiv preprint arXiv:2408.01439}, 2024.

\bibitem{martyn2025parallel}
J.~M. Martyn, Z.~M. Rossi, K.~Z. Cheng, Y.~Liu, and I.~L. Chuang.
\newblock Parallel quantum signal processing via polynomial factorization.
\newblock {\em Quantum}, 9:1834, 2025.

\bibitem{MartynRossiTanEtAl2021}
J.~M. Martyn, Z.~M. Rossi, A.~K. Tan, and I.~L. Chuang.
\newblock Grand unification of quantum algorithms.
\newblock {\em PRX Quantum}, 2:040203, 2021.

\bibitem{MotlaghWiebe2024}
D.~Motlagh and N.~Wiebe.
\newblock Generalized quantum signal processing.
\newblock {\em PRX Quantum}, 5:020368, 2024.

\bibitem{nemeth2023variants}
B.~N{\'e}meth, B.~K{\"o}v{\'e}r, B.~Kulcs{\'a}r, R.~B. Mikl{\'o}si, and
  A.~Gily{\'e}n.
\newblock On variants of multivariate quantum signal processing and their
  characterizations.
\newblock {\em arXiv preprint arXiv:2312.09072}, 2023.

\bibitem{ni2025inverse}
H.~Ni, R.~Sarkar, L.~Ying, and L.~Lin.
\newblock Inverse nonlinear fast {F}ourier transform on {SU(2)} with
  applications to quantum signal processing.
\newblock {\em arXiv preprint arXiv:2505.12615}, 2025.

\bibitem{ni2024fast}
H.~Ni and L.~Ying.
\newblock Fast phase factor finding for quantum signal processing.
\newblock {\em arXiv preprint arXiv:2410.06409}, 2024.

\bibitem{paige1994history}
C.~C. Paige and M.~Wei.
\newblock History and generality of the {CS} decomposition.
\newblock {\em Linear Algebra Appl.}, 208:303--326, 1994.

\bibitem{perez2021one}
A.~P{\'e}rez-Salinas, D.~L{\'o}pez-N{\'u}{\~n}ez, A.~Garc{\'\i}a-S{\'a}ez,
  P.~Forn-D{\'\i}az, and J.~I. Latorre.
\newblock One qubit as a universal approximant.
\newblock {\em Physical Review A}, 104(1):012405, 2021.

\bibitem{RossiBastidasMunroEtAl2023}
Z.~M. Rossi, V.~M. Bastidas, W.~J. Munro, and I.~L. Chuang.
\newblock Quantum signal processing with continuous variables.
\newblock {\em arXiv preprint arXiv:2304.14383}, 2023.

\bibitem{RossiChuang2022}
Z.~M. Rossi and I.~L. Chuang.
\newblock Multivariable quantum signal processing ({M-QSP}): prophecies of the
  two-headed oracle.
\newblock {\em Quantum}, 6:811, 2022.

\bibitem{schur1918potenzreihen}
J.~Schur.
\newblock {\"U}ber potenzreihen, die im innern des einheitskreises
  beschr{\"a}nkt sind.
\newblock {\em Journal f{\"u}r die reine und angewandte Mathematik (Crelles
  Journal)}, pages 122--145, 1918.

\bibitem{simon2004canonical}
B.~Simon.
\newblock A canonical factorization for meromorphic {H}erglotz functions on the
  unit disk and sum rules for {J}acobi matrices.
\newblock {\em J. Funct. Anal.}, 214:396--409, 2004.

\bibitem{singh2025towards}
S.~Singh, B.~Royer, and S.~M. Girvin.
\newblock Towards non-abelian quantum signal processing: Efficient control of
  hybrid continuous-and discrete-variable architectures.
\newblock {\em arXiv preprint arXiv:2504.19992}, 2025.

\bibitem{sunderhauf2023generalized}
C.~S{\"u}nderhauf.
\newblock Generalized quantum singular value transformation.
\newblock {\em arXiv preprint arXiv:2312.00723}, 2023.

\bibitem{szego1939ortho}
G.~Szeg{\H o}.
\newblock {\em Orthogonal Polynomials}, volume~23.
\newblock American Mathematical Society, 1939.

\bibitem{tanaka1972some}
S.~Tanaka.
\newblock Some remarks on the modified {K}orteweg-de {V}ries equations.
\newblock {\em Publications of the Research Institute for Mathematical
  Sciences}, 8:429--437, 1972.

\bibitem{tang2024cs}
E.~Tang and K.~Tian.
\newblock A {CS} guide to the quantum singular value transformation.
\newblock In {\em 2024 Symposium on Simplicity in Algorithms (SOSA)}, pages
  121--143. SIAM, 2024.

\bibitem{tao2012nonlinear}
T.~Tao and C.~Thiele.
\newblock Nonlinear {F}ourier analysis.
\newblock {\em arXiv preprint arXiv:1201.5129}, 2012.

\bibitem{tsai2005nlft}
Y.-J. Tsai.
\newblock {\em {SU}(2) nonlinear {F}ourier transform}.
\newblock PhD thesis, University of California, Los Angeles, 2005.

\bibitem{volberg2002inverse}
A.~Volberg and P.~Yuditskii.
\newblock On the inverse scattering problem for {J}acobi matrices with the
  spectrum on an interval, a finite system of intervals or a {C}antor set of
  positive length.
\newblock {\em Commun. Math. Phys.}, 226:567--605, 2002.

\bibitem{WangDongLin2022}
J.~Wang, Y.~Dong, and L.~Lin.
\newblock On the energy landscape of symmetric quantum signal processing.
\newblock {\em Quantum}, 6:850, 2022.

\bibitem{weiss1962derivation}
M.~Weiss and G.~L. Weiss.
\newblock A derivation of the main results of the theory of {$H^p$} spaces.
\newblock {\em Revista de la Union Matematica Argentina}, 20:63--71, 1962.

\bibitem{winebrenner1998linear}
D.~P. Winebrenner and J.~Sylvester.
\newblock Linear and nonlinear inverse scattering.
\newblock {\em SIAM J. Appl. Math.}, 59:669--699, 1998.

\bibitem{Ying2022}
L.~Ying.
\newblock Stable factorization for phase factors of quantum signal processing.
\newblock {\em Quantum}, 6:842, 2022.

\end{thebibliography}

\end{document}